\documentclass[aps,prl,reprint,superscriptaddress,longbibliography,nofootinbib,floatfix]{revtex4-2}
\usepackage{amsmath,amssymb,mathtools,bm,graphicx,mathrsfs}
\usepackage[colorlinks=true,linkcolor=blue,citecolor=blue,urlcolor=blue]{hyperref}
\newcommand{\dd}{\mathrm{d}}
\newcommand{\ii}{\mathrm{i}}
\newcommand{\eps}{\varepsilon}
\newcommand{\BZ}{\mathrm{BZ}}
\newcommand{\BerryA}{\mathscr{A}}
\newcommand{\BerryF}{\mathscr{F}}

\graphicspath{{./}}

\newcommand{\ee}{\mathrm{e}}
\newcommand{\Tr}{\operatorname{Tr}}
\newcommand{\Bc}{\mathcal B}
\newcommand{\CB}{C_{\Bc}}
\newcommand{\rhoA}{\rho_{\mathcal A}}
\newcommand{\jA}{j_{\mathcal A}}
\newcommand{\NA}{\mathcal N_{\mathcal A}}
\newcommand{\tA}{\bar t_{\mathcal A}}
\newcommand{\PA}{P_{\mathcal A}}
\newcommand{\kpair}[2]{\langle\!\langle #1,#2\rangle\!\rangle_{\mathcal A}}

\begin{document}

\title{Thouless Pumping of Wave Action in Photonic Time Crystals}
\author{Minwook Kyung}
\author{Younsung Kim}
\author{Kyungmin Lee}
\affiliation{Department of Physics, Korea Advanced Institute of Science and Technology (KAIST), Daejeon 34141, Republic of Korea}
\author{Hee Chul Park}
\affiliation{Department of Physics, Pukyong National University, Busan 48513, Republic of Korea}
\author{Moon Jip Park}
\affiliation{Department of Physics, Hanyang University, Seoul 04763, Republic of Korea}
\author{Bumki Min}
\email{bmin@kaist.ac.kr}
\affiliation{Department of Physics, Korea Advanced Institute of Science and Technology (KAIST), Daejeon 34141, Republic of Korea}
\date{\today}

\begin{abstract}
A topological pump transports a conserved quantity. Yet temporal modulation exchanges energy with the field, so the transported quantity must be identified from Maxwell's equations. Here we connect band topology to the temporal displacement of conserved Maxwell wave action in a photonic time crystal. The temporal Wannier function of an isolated positive-action band carries unit action with zero net energy flux, and its temporal center is a Berry polarization. Over one adiabatic cycle, that center shifts by the band Chern number: one cell for a rigid translation and two for a two-tone cycle within a single band.
\end{abstract}

\maketitle

A Thouless pump turns the geometry of a periodic band into the quantized transport of particles: each adiabatic cycle carries an integer number across the crystal, set by a band Chern number \cite{Thouless1983,Simon1983,Nakajima2016,Braver2022,Kraus2012,Zilberberg2018}. In a Thouless pump, conservation identifies the quantity whose transport is quantized. In the photonic realizations, stationary propagation conserves a positive power norm~\cite{Kraus2012,Zilberberg2018}. The transported density is proportional to the optical intensity, and its normalized first moment defines the transverse beam center. Electromagnetic field energy does not meet that condition in a photonic time crystal (PTC)~\cite{Galiffi2022,ParkFloquet2022,LiberalMomentum,Asgari2024,ParkSE2025,Bae2026,KimQPTC2026}: periodic modulation of the permittivity exchanges energy with the field, and the field-energy balance acquires a source term~\cite{LeeEnergyTransport}. The temporal counterpart, the time-integrated longitudinal energy flux, is not generally conserved along $z$ and vanishes for the action-normalized full-zone Wannier state~\cite{SM}. The task is therefore to identify a conserved Maxwell density whose temporal center can be related to the band geometry.

Band topology and wave-action conservation provide complementary descriptions of time-varying photonic media. An early study defined Zak phases for PTC bands and predicted a localized state at a temporal interface \cite{Lustig2018}. Subsequent work extended this picture to chiral-symmetric, quasiperiodic, non-Hermitian, and space-time photonic systems~\cite{YangGarciaVidal2025,NiAlu2025,Xu2025,Jiang2025,CaballeroHuidobro2026}; temporal interface states were also observed~\cite{Xiong2025,Ren2025}. In parallel, conservation studies identified quantities preserved under temporal modulation: a photon count that assigns negative weight to negative-frequency quanta is conserved in space-time gratings while the energy grows~\cite{PendryOptica2022,Pendry2023,Zhang2024}, and conservation of wave action renders the Floquet scattering matrix of a finite time-modulated medium pseudounitary~\cite{Globosits2024}. Temporal Wannier pumping has also been studied in driven quantum lattices~\cite{Braver2022}, and Bogoliubov Wannier pumping in driven-dissipative Kerr arrays~\cite{Ravets2025}. For Maxwell waves, the key question is how band topology controls the temporal displacement of an exactly conserved local density.

Here we show that a photonic time crystal supports quantized temporal transport of conserved Maxwell wave action. For an isolated positive-action band, the action-normalized Wannier Cauchy state carries unit action with zero net energy flux, and its temporal center has an exact Berry-phase representation. Maxwell propagation approaches the corresponding Chern-quantized displacement in the adiabatic limit, even when the physical wavepacket broadens.

\begin{figure}[htb!]
\centering
\includegraphics[width=1.00\columnwidth]{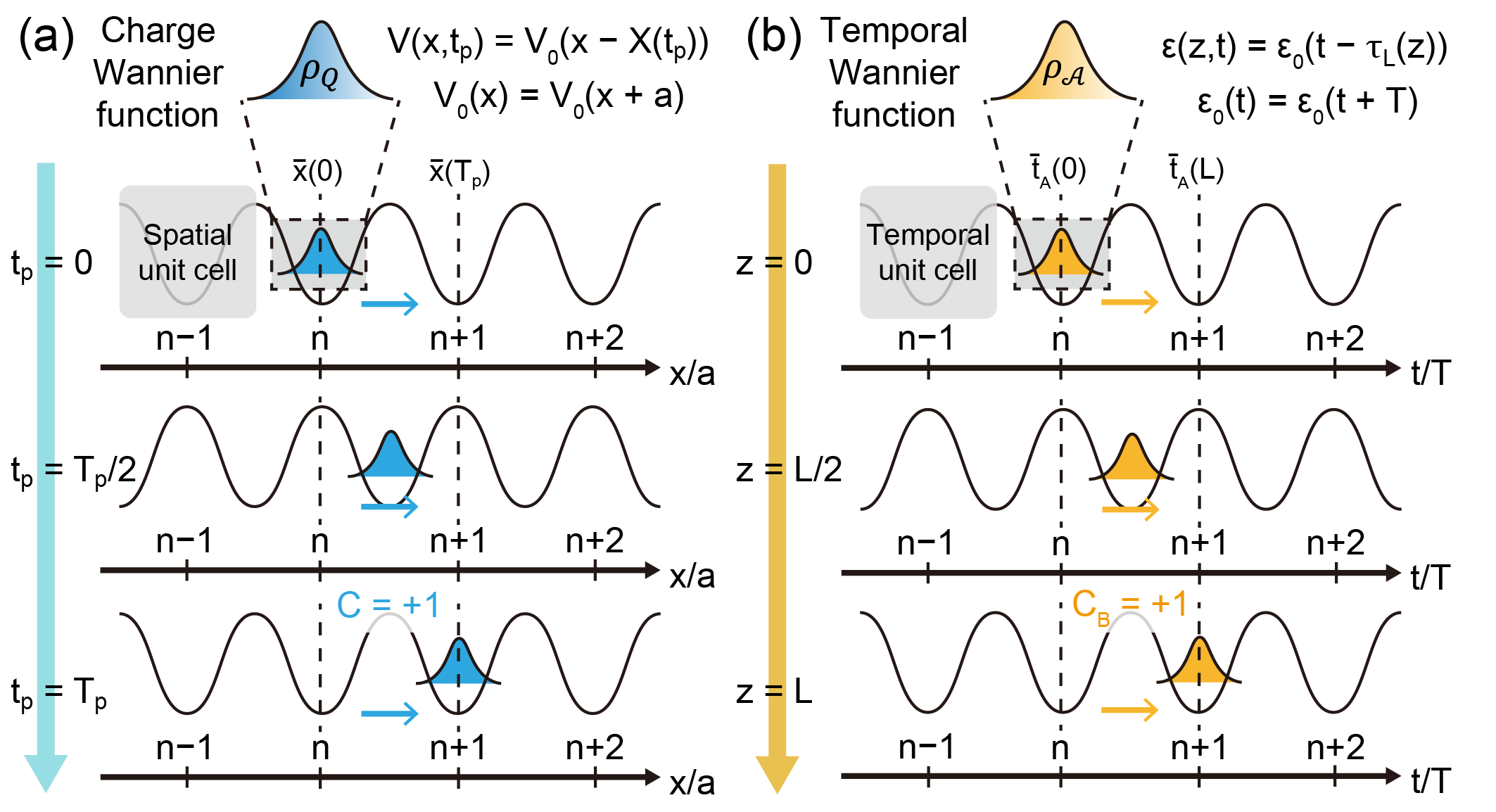}
\caption{\textbf{Charge and wave-action pumping.} (a) A rigidly sliding lattice potential carries a localized Wannier function with charge density $\rho_Q$. Its center $\bar x$ shifts by $C$ spatial cells per pump period. (b) A permittivity sliding along $z$ carries a temporal Wannier function with action density $\rho_{\mathcal A}$. Its center $\bar t_{\mathcal A}$ advances by $C_{\mathcal B}$ temporal cells across the device. The sketches show the elementary $C=C_{\mathcal B}=+1$ case; the profiles, and their placement relative to the modulation, are schematic.}
\label{fig:concept}
\end{figure}

\begin{figure}[htb!]
\centering
\includegraphics[width=1.00\columnwidth]{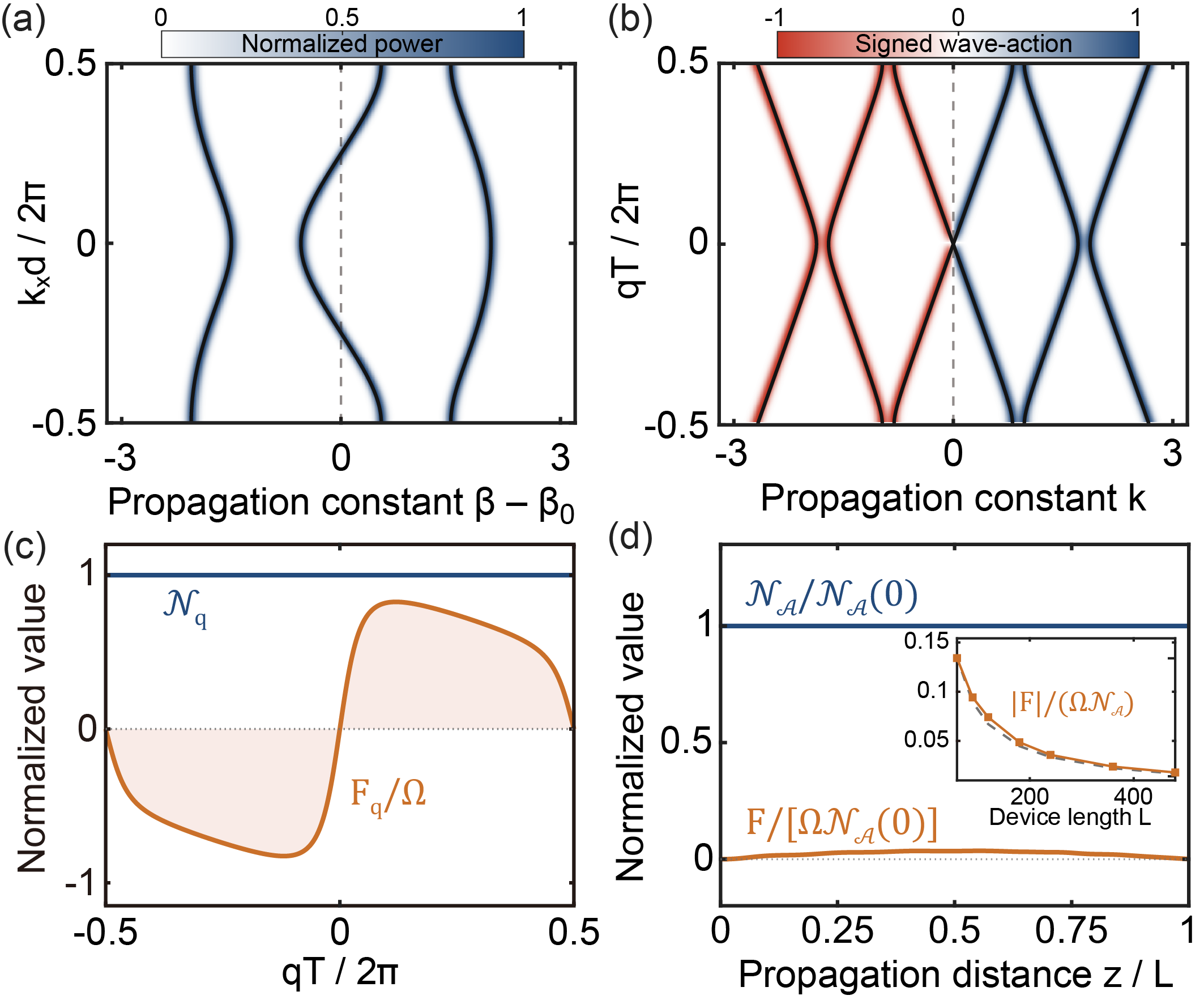}
\caption{\textbf{Positive norm, signed action, and vanishing net flux.} (a) Waveguide pump~\cite{Kraus2012,Zilberberg2018}: paraxial evolution is first order in $z$ and conserves a positive power norm in the forward-propagating sector. (b) Temporal Bloch spectrum: modes come in $\pm k$ pairs at each quasifrequency $q$, carrying opposite action signs (blue positive, red negative); a band must be selected. (c) Per-state wave action $\mathcal N_q$ (blue) and time-integrated energy flux per unit action $F_q$ (orange, shaded; units of $\Omega$) of the selected band; $F_q$ is odd in $q$, so its zone average vanishes. (d) Wave action $\mathcal N_{\mathcal A}$ and time-integrated energy flux $F$ along a device of length $L=240\,c/\Omega$; inset, the largest action-weighted mean frequency $|F|/(\Omega\mathcal N_{\mathcal A})$ for $L=60$ to $480\,c/\Omega$. Panels (b)--(d) use the rigid-pump medium of Eq.~\eqref{eq:harmonic-medium} [\hyperref[fig:c1]{Fig.~\ref*{fig:c1}}], at $\tau=0$ in (b),(c) and over one complete rigid cycle in (d); parameters are listed in the Supplemental Material~\cite{SM}.}
\label{fig:eflux}
\end{figure}

The modulation profile varies along $z$ through a cycle parameter $\tau$, $\eps(z,t)=\eps[t;\tau(z)]$, while the periodic time $t$ defines the temporal unit cell. The continuity equation yields a conserved integral over $t$, whose normalized first moment defines the action polarization. Over one closed cycle it advances by the band Chern number: $C_{\mathcal B}=1$ for a rigid translation, intrinsic $C_{\mathcal B}=2$ for a primitive two-tone cycle in a single band. \hyperref[fig:concept]{Figure~\ref*{fig:concept}} summarizes the correspondence with the Thouless charge pump. The PTC differs from the waveguide pump through its opposite-action branches and vanishing net energy flux [\hyperref[fig:eflux]{Fig.~\ref*{fig:eflux}}].

\emph{Exact wave-action conservation.---}
We consider a lossless, nondispersive dielectric with relative permittivity $\eps(z,t)$. In the Coulomb gauge for a single transverse polarization, the vector potential $A\equiv A_x$ satisfies $E_x=-\partial_tA$ and $H_y=\mu^{-1}\partial_zA$, and obeys
\begin{equation}
\partial_t\!\left[\eps(z,t)\partial_tA\right]-\mu^{-1}\partial_z^2A=0,
\qquad
\eps(z,t)>0,
\label{eq:maxwell}
\end{equation}
with constant $\mu>0$ and $c=1$. We use the complex representation $A=A_1+\ii A_2$, where $A_1$ and $A_2$ are two real quadrature solutions of the same Maxwell equation~\cite{SM}. The equation is invariant under the global phase rotation $A\rightarrow e^{\ii\theta}A$. The associated bilinear current has components $\rho_{\mathcal A}$ and $j_{\mathcal A}$~\cite{Hayes1970,Globosits2024},
\begin{equation}
\begin{aligned}
\rho_{\mathcal A}
&=\frac{A^*\partial_zA-A\partial_zA^*}{2\ii\mu},\\
j_{\mathcal A}
&=-\frac{\eps}{2\ii}\left(A^*\partial_tA-A\partial_tA^*\right).
\end{aligned}
\label{eq:action-components}
\end{equation}
These components obey the exact continuity equation
\begin{equation}
\partial_z\rho_{\mathcal A}+\partial_tj_{\mathcal A}=0.
\label{eq:action-current}
\end{equation}
The electromagnetic-energy balance contains the source term $-(\partial_t\eps)|\partial_tA|^2/2$, so temporal modulation exchanges energy with the field \cite{LeeEnergyTransport}, while the wave-action current remains exactly divergence-free.

The wave action of the selected band adds across the quasifrequency zone, and the energy flux cancels. On the selected positive-action branch, the cell-integrated harmonic action contributions have one sign, whereas the energy-flux contributions are weighted by their signed physical frequencies~\cite{SM,PendryOptica2022,Pendry2023}. Real-valuedness of the permittivity makes the energy flux per unit action of a temporal Bloch state [\hyperref[fig:eflux]{Fig.~\ref*{fig:eflux}(b)}] odd in the quasifrequency, $F_{-q}=-F_q$, so its zone average vanishes identically [\hyperref[fig:eflux]{Fig.~\ref*{fig:eflux}(c)}]. The temporal Wannier function of an isolated positive-action band therefore carries unit wave action together with zero net energy flux. Finite-length propagation leaves the small residual energy flux shown in \hyperref[fig:eflux]{Fig.~\ref*{fig:eflux}(d)}.

For a packet localized in $t$, the conserved wave action and the center of its temporal distribution are
\begin{equation}
\mathcal N_{\mathcal A}=\int_{-\infty}^{+\infty}\rho_{\mathcal A}\,\dd t,
\qquad
\bar t_{\mathcal A}=\frac{1}{\mathcal N_{\mathcal A}}\int_{-\infty}^{+\infty}t\rho_{\mathcal A}\,\dd t.
\end{equation}
Equation~\eqref{eq:action-current} then gives
\begin{equation}
\frac{\dd\bar t_{\mathcal A}}{\dd z}
=\frac{1}{\mathcal N_{\mathcal A}}\int_{-\infty}^{+\infty}j_{\mathcal A}\,\dd t,
\label{eq:transport}
\end{equation}
so the action center advances at a rate set by the net action current.

\begin{figure}[htb!]
\centering
\includegraphics[width=1.00\columnwidth]{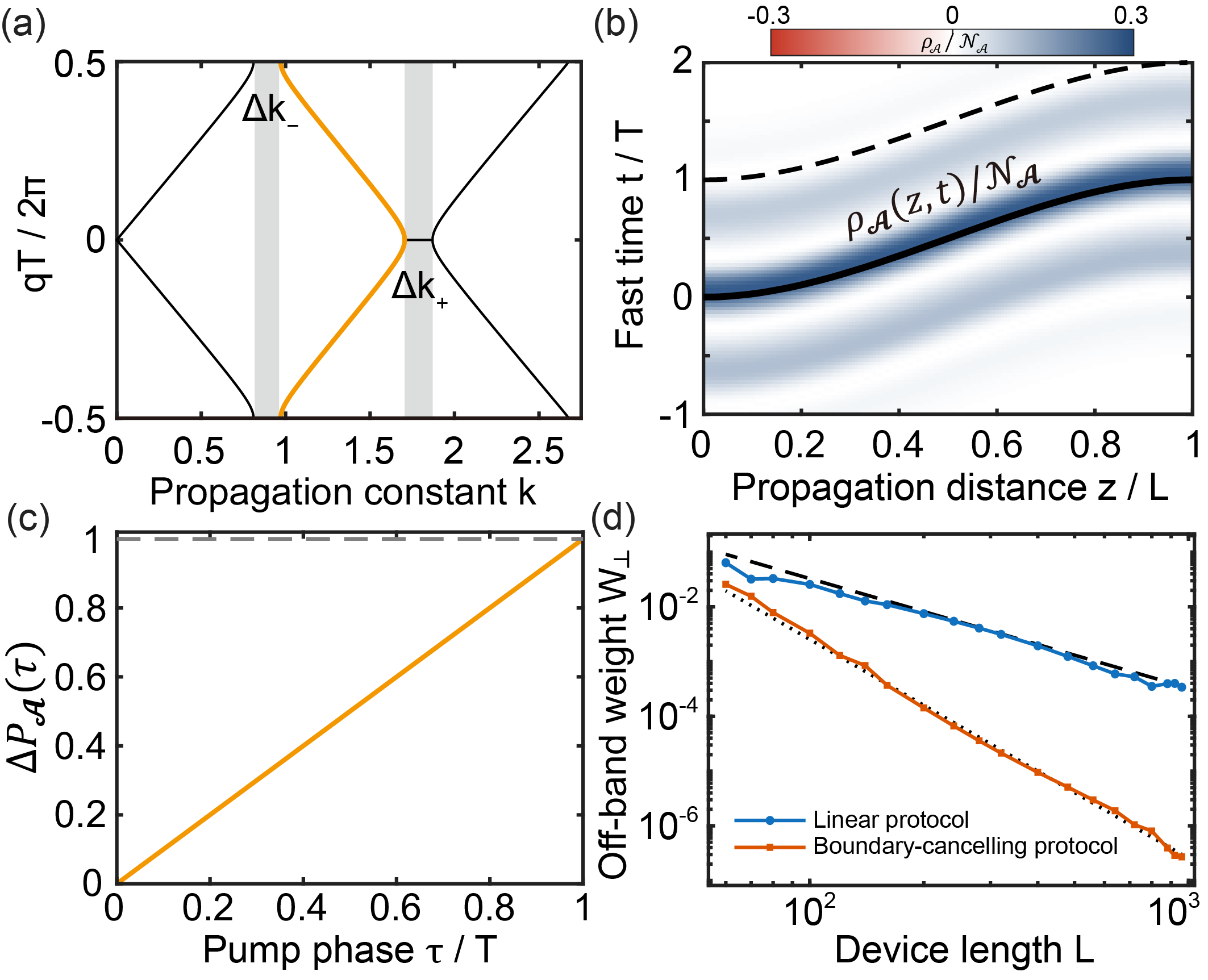}
\caption{\textbf{One-cell wave-action pump.} (a) The selected forward band (orange) remains isolated by the propagation-constant gaps $\Delta k_-$ and $\Delta k_+$. (b) Maxwell propagation over one rigid translation, showing $\rho_{\mathcal A}(z,t)/\mathcal N_{\mathcal A}$ in the smooth rephased frame~\cite{SM}; the action center and its next-cell image (dashed) advance one temporal cell across the device. (c) The action polarization winds once around the temporal cell, giving $C_{\mathcal B}=+1$. (d) The off-band weight $W_\perp(L)$ is consistent with $L^{-4}$ when the pump starts and ends at rest (red), and with $L^{-2}$ at finite starting rate (blue)~\cite{SM}. Parameters are listed in the Supplemental Material~\cite{SM}.}
\label{fig:c1}
\end{figure}

\emph{The action center as a Berry polarization.---}
The permittivity $\eps(t;\tau)$ remains real and positive and satisfies
\begin{equation}
\eps(t+T;\tau)=\eps(t;\tau),
\qquad
\eps(t;\tau+T)=\eps(t;\tau).
\label{eq:periodic-medium}
\end{equation}
Here $T$ is the common period and $\Omega=2\pi/T$ the modulation frequency, while $\tau\in[0,T)$ parametrizes one closed modulation cycle that a device of length $L$ traverses once through $\eps(z,t)=\eps[t;\tau_L(z)]$. We seek solutions $A(z,t)=e^{\ii k_m(q,\tau)z}\psi_{mq\tau}(t)$ with propagation constant $k_m(q,\tau)$. At fixed $\tau$, the temporal factor is a Bloch mode of band index $m$ and quasifrequency $q\in\BZ\equiv[-\pi/T,\pi/T)$, written as $\psi_{mq\tau}(t)=e^{\ii qt}u_{mq\tau}(t)$ with the periodic factor $u_{mq\tau}(t+T)=u_{mq\tau}(t)$. Substituting into Eq.~\eqref{eq:maxwell} at fixed $\tau$ gives an eigenproblem for the periodic factor, with $\mathcal L_q[\tau]\equiv-\mu\left(\partial_t+\ii q\right)\eps(t;\tau)\left(\partial_t+\ii q\right)$,
\begin{equation}
\mathcal L_q[\tau]\,u_{mq\tau}=k_m^2(q,\tau)\,u_{mq\tau}.
\label{eq:band-eigenproblem}
\end{equation}
The operator $\mathcal L_q[\tau]$ is self-adjoint, and positivity of $\eps$ implies $k_m^2(q,\tau)\ge0$~\cite{SM}. The branch with $k_m>0$ carries positive wave action, whereas its partner at $-k_m$ carries the opposite sign~\cite{SM} [\hyperref[fig:eflux]{Fig.~\ref*{fig:eflux}(b)}]. We select a forward band $\mathcal B$ that remains isolated from every other forward and backward branch and remains separated from $k=0$ throughout the cycle.

We normalize the periodic factor over one cell, $\int_0^T|u_{\mathcal Bq\tau}|^2\,\dd t=1$, and rescale the full mode by $\sqrt{\mu/k_{\mathcal B}}$ to unit action. We then construct the action-normalized temporal Wannier function by occupying the full temporal Brillouin zone of the band with uniform weight~\cite{Kohn1959}. Together, this function and its branch-matched $z$ derivative form the Wannier Cauchy state, the pair of initial data required by the second-order Maxwell equation~\cite{SM}. In a $q$-periodic gauge chosen smoothly along the cycle, with the end frames related by a transition phase, we define the temporal action polarization by
\begin{equation}
\begin{aligned}
P_{\mathcal A}(\tau)
&\equiv\frac{\bar t_{\mathcal A}(\tau)}{T}\pmod{1},\\
&=\frac{1}{2\pi}\int_{\BZ}\BerryA_q(q,\tau)\,\dd q,\\
\BerryA_q(q,\tau)
&=\ii\int_0^T u_{\mathcal B q\tau}^*(t)\partial_q u_{\mathcal B q\tau}(t)\,\dd t.
\end{aligned}
\label{eq:polarization}
\end{equation}
The normalized first moment of the conserved wave-action density is exactly the Berry polarization of the selected band \cite{Berry1984,KingSmith1993,Resta1994}; the $q$-dependent factor $\sqrt{\mu/k_{\mathcal B}}$ contributes no additional Berry connection~\cite{SM}.

Because $P_{\mathcal A}$ is defined modulo one temporal cell, the pumped quantity $\Delta P_{\mathcal A}$ is the net change of a continuous branch of $P_{\mathcal A}(\tau)$ followed along the cycle. The change of this continuously unwrapped polarization gives the action-center displacement in units of $T$. The pump-direction Berry connection $\BerryA_\tau$ is the connection of Eq.~\eqref{eq:polarization} with $\partial_q$ replaced by $\partial_\tau$. Over one closed pump cycle,
\begin{equation}
\begin{aligned}
\Delta P_{\mathcal A}
&=\frac{1}{2\pi}\int_0^T\dd\tau\int_{\BZ}\dd q\,\BerryF_{\tau q}\\
&=C_{\mathcal B}\in\mathbb Z,\\
\BerryF_{\tau q}
&=\partial_\tau\BerryA_q-\partial_q\BerryA_\tau.
\end{aligned}
\label{eq:chern}
\end{equation}
At fixed harmonic truncation, if the pump parameter advances through one complete cycle across the device, with the pump switched on and off along a sufficiently smooth protocol~\cite{SM} so that it starts and ends at rest, $\tau_L'(0)=\tau_L'(L)=0$, then the field launched by the Wannier Cauchy state satisfies~\cite{SM}
\begin{equation}
\frac{\bar t_{\mathcal A}(L)-\bar t_{\mathcal A}(0)}{T}=C_{\mathcal B}+O(L^{-1}).
\label{eq:adiabatic}
\end{equation}
The positive modal weight outside the selected instantaneous forward band, $W_\perp(L)$, obeys $W_\perp=O(L^{-2})$ at finite starting rate, improving to $O(L^{-4})$ when the pump starts and ends at rest~\cite{SM}. In the adiabatic limit the same integer follows in the instantaneous band frame as the winding of the action polarization, computed by a Wilson loop over the quasifrequency zone~\cite{Fukui2005,SM}. The action center is reconstructed from phase-resolved $E_x$ and $H_y$, whereas the Wilson loop uses selected-band eigenvector overlaps.

\begin{figure}[htb!]
\centering
\includegraphics[width=1.00\columnwidth]{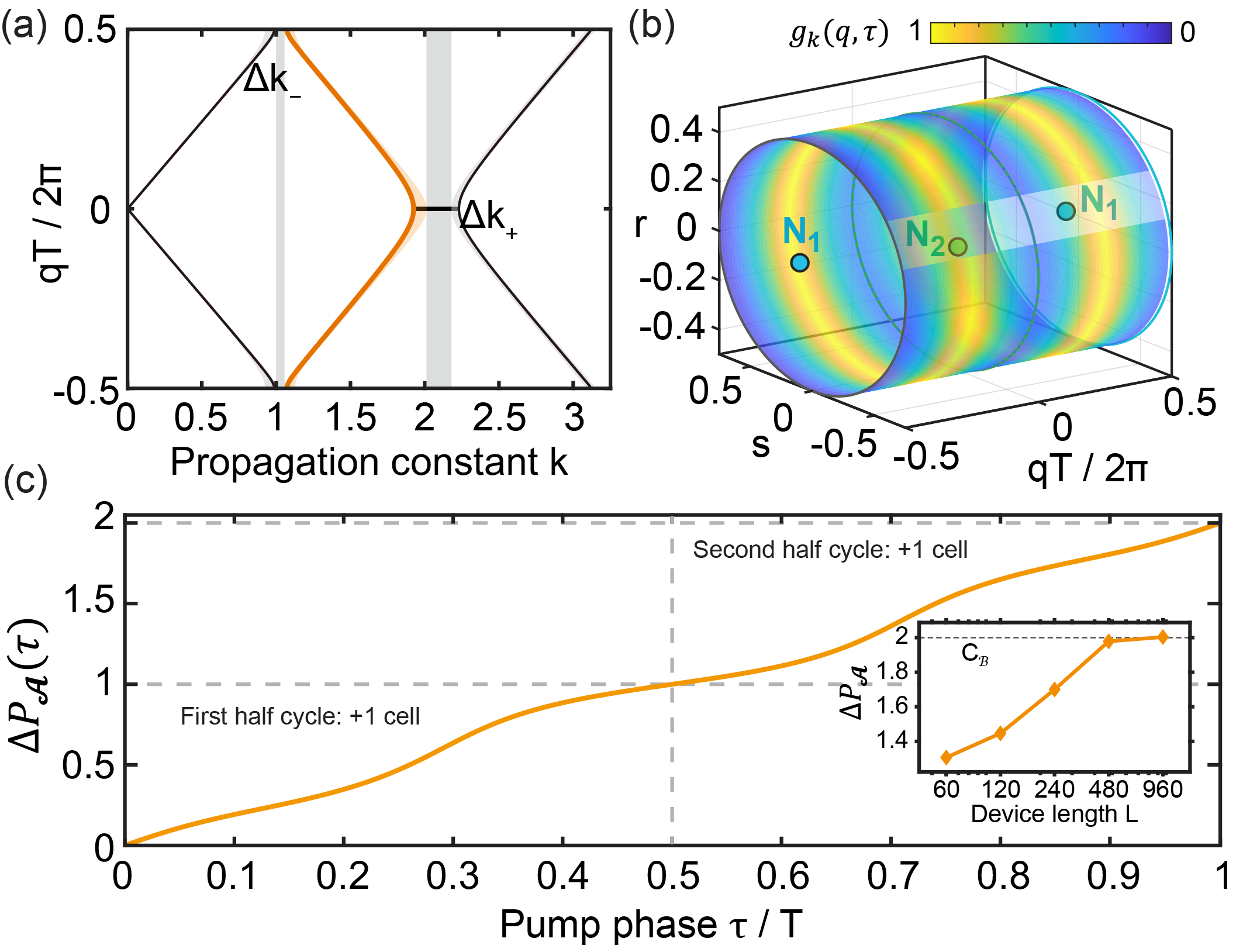}
\caption{\textbf{Intrinsic two-cell pump in a single band.} (a) The target forward band (orange) remains isolated throughout the two-tone cycle by the protecting gaps (gray). (b) The solid pump domain traced out by the closed control disk across the quasifrequency zone, colored by the protecting gap $g_k(q,\tau)$, the minimum distance in $k$ between $k_{\mathcal B}$ and every other signed branch. It contains $\mathsf N_2$ at $q=0$ (upper neighbor) and $\mathsf N_1$ at the zone edge (lower neighbor), drawn at both edges because $q$ and $q+\Omega$ are identified; each contributes $+1$ for the chosen loop orientation. (c) The action polarization advances by one temporal cell in each half cycle, giving $C_{\mathcal B}=+2$. Inset: the action-center displacement $\Delta\bar t_{\mathcal A}/T$ from quasifrequency-resolved propagation of Eq.~\eqref{eq:maxwell} through devices of length $L=60$ to $960\,c/\Omega$, approaching the quantized value of two cells. Parameters are listed in the Supplemental Material~\cite{SM}.}
\label{fig:c2}
\end{figure}

\emph{One-cell pump.---}
To realize a nonzero invariant, we take the simplest cycle, a rigid translation of the modulation profile through one period. Write the two-harmonic medium as
\begin{equation}
\eps(t;\tau)=\eps_b+2\,\mathrm{Re}\!\left[\sum_{p=1,2}\hat\eps_p(\tau)\,e^{\ii p\Omega t}\right],
\label{eq:harmonic-medium}
\end{equation}
with $\hat\eps_p(\tau)$ the complex amplitude of the $p$th temporal harmonic. The cycle is $\eps_{\mathrm r}(t;\tau)=\eps_0(t-\tau)$, for which each harmonic acquires the phase of one common time shift, $\hat\eps_p(\tau)=\hat\eps_p(0)\,e^{-\ii p\Omega\tau}$, and the profile slides without changing shape. The instantaneous Wannier profile therefore translates by one temporal cell as the medium's clock advances by one period. At the end of the cycle the chosen Bloch frame satisfies $u(q,T,t)=e^{-\ii qT}u(q,0,t)$, so the cycle-end transition function $\chi(q)=e^{-\ii qT}$ has winding $-1$. With the convention of Eq.~\eqref{eq:chern} the selected band therefore carries $C_{\mathcal B}=-\operatorname{wind}\chi=+1$.

\hyperref[fig:c1]{Figure~\ref*{fig:c1}} verifies the resulting transport through quasifrequency-resolved propagation of Eq.~\eqref{eq:maxwell}. For the second forward band, which remains isolated throughout the cycle [\hyperref[fig:c1]{Fig.~\ref*{fig:c1}(a)}], finite-length propagation of the Wannier Cauchy state advances the action center by one temporal cell [\hyperref[fig:c1]{Fig.~\ref*{fig:c1}(b)}], and the ideal action polarization winds once around the cell to the same integer [\hyperref[fig:c1]{Fig.~\ref*{fig:c1}(c)}]. The off-band weight quantifies nonadiabatic mode conversion out of the selected band. It falls as $L^{-4}$ once the pump starts and ends at rest and as $L^{-2}$ at finite starting rate [\hyperref[fig:c1]{Fig.~\ref*{fig:c1}(d)}].

\emph{Intrinsic two-cell pump in one band.---}
The rigid cycle reaches only $C_{\mathcal B}=1$, yet the Chern number of an isolated band is not restricted to unity. To realize $C_{\mathcal B}=2$, we vary the two temporal harmonics independently over the same pump period $\tau\in[0,T)$. The amplitudes $\hat\eps_1(\tau)$ and $\hat\eps_2(\tau)$ of Eq.~\eqref{eq:harmonic-medium} trace the primitive control loop $(s,r)=(R_s\cos\Omega\tau,-R_r\sin\Omega\tau)$, an ellipse of semi-axes $R_s$ and $R_r$ in the cosine and sine quadratures of the two harmonics.

The loop encloses a disk, and sweeping it over the quasifrequency zone gives a solid domain that contains two symmetry-enforced conical nodes in the three-parameter space $(q,s,r)$~\cite{SM}: at one node the target band meets its lower neighbor, and at the other it meets its upper neighbor. Although the linearized crossings have opposite Jacobian signs, the target band lies above one crossing and below the other. These two sign reversals compensate, so each node contributes $+1$ for the chosen loop orientation and the selected band acquires $C_{\mathcal B}=2$. The target band remains isolated throughout the two-tone cycle, and the gap protecting it remains open everywhere on its boundary torus [\hyperref[fig:c2]{Figs.~\ref*{fig:c2}(a)} and \hyperref[fig:c2]{\ref*{fig:c2}(b)}]. The continuously unwrapped action polarization advances by two temporal cells [\hyperref[fig:c2]{Fig.~\ref*{fig:c2}(c)}], and quasifrequency-resolved propagation of Eq.~\eqref{eq:maxwell} approaches the same displacement at finite length [\hyperref[fig:c2]{Fig.~\ref*{fig:c2}(c)}, inset]. In the adiabatic limit, a single primitive cycle of one band thus pumps two temporal cells.

\emph{Discussion.---}
A photonic time crystal does not conserve electromagnetic field energy: the modulation feeds energy to the field and amplifies waves within the momentum gaps. The conserved quantity is wave action, whose temporal center advances by the band Chern number in the adiabatic pump. The wave-action continuity equation, Eq.~\eqref{eq:action-current}, follows directly from Maxwell's equations and remains exact for exponentially growing solutions in the momentum gaps. The positive energy-density center follows source-driven dynamics and generally differs from the band Berry polarization~\cite{SM,LeeEnergyTransport}.

Causal dispersion requires an enlarged field--matter formulation, while a microscopic treatment of material loss additionally introduces reservoir degrees of freedom~\cite{HooperCapers2025,Bakunov2021,Bakunov2024,HorsleyBaker2025}. With dispersion, the dynamical state combines field and matter degrees of freedom~\cite{Serra2024}, and the relevant action is carried jointly by the two sectors. The nondispersive model studied here therefore supports an exact field-only wave-action current. The resulting transport is a classical Chern pump of wave action. Extending this framework to dispersive media requires the complete polaritonic spectrum at temporal interfaces.

\section*{Acknowledgments}
This work was supported by the National Research Foundation of Korea (NRF), funded by the Korean government (\mbox{RS-2022-NR070636}), and by the Samsung Science and Technology Foundation (\mbox{SSTF-BA240202}). M.J.P. acknowledges support from the NRF grant \mbox{RS-2025-25464760}, and H.C.P. from the NRF grant \mbox{RS-2023-00278511}, both funded by the Korean government.

%

\clearpage
\onecolumngrid

\setlength{\parskip}{0pt}
\setlength{\parindent}{1em}
\allowdisplaybreaks[1]

\setcounter{section}{0}
\setcounter{subsection}{0}
\setcounter{equation}{0}
\setcounter{figure}{0}
\setcounter{table}{0}
\renewcommand{\theequation}{S\arabic{equation}}
\renewcommand{\theHequation}{sm.\arabic{equation}}
\renewcommand{\thefigure}{S\arabic{figure}}
\renewcommand{\thetable}{S\arabic{table}}
\providecommand{\theHfigure}{}
\providecommand{\theHtable}{}
\providecommand{\theHsection}{}
\renewcommand{\theHfigure}{sm.fig.\arabic{figure}}
\renewcommand{\theHtable}{sm.tab.\arabic{table}}
\renewcommand{\theHsection}{sm.sec.\arabic{section}}
\setcounter{secnumdepth}{3}

\begin{center}
{\large\bfseries Supplemental Material for ``Thouless Pumping of Wave Action in Photonic Time Crystals''\par}
\vspace{0.8em}
Minwook Kyung,$^{1}$ Younsung Kim,$^{1}$ Kyungmin Lee,$^{1}$ Hee Chul Park,$^{2}$ Moon Jip Park,$^{3}$ and Bumki Min$^{1,*}$\\[0.35em]
$^{1}$Department of Physics, Korea Advanced Institute of Science and Technology (KAIST), Daejeon 34141, Republic of Korea\\
$^{2}$Department of Physics, Pukyong National University, Busan 48513, Republic of Korea\\
$^{3}$Department of Physics, Hanyang University, Seoul 04763, Republic of Korea\\[0.35em]
(Dated: \today)\\
$^{*}$bmin@kaist.ac.kr
\end{center}
\vspace{0.5em}

\section{Exact wave-action current}
We take one transverse polarization, with $\eps(z,t)$ real and finite and $\mu$ real, constant, and nonzero. With $E_x=-\partial_tA$ and $H_y=\mu^{-1}\partial_zA$, Maxwell's equations reduce to
\begin{equation}
\label{eq:sm:maxwell}
 \partial_t(\eps\partial_tA)-\mu^{-1}\partial_z^2A=0.
\end{equation}
We fix the residual constant shift of $A$. The complex field $A=A_1+\ii A_2$ combines two real solutions of \eqref{eq:sm:maxwell} that form a quadrature pair about a common phase origin. Invariance under $A\mapsto\ee^{\ii\alpha}A$ gives the bilinear conserved current~\cite{SMHayes1970,SMGlobosits2024}
\begin{equation}
\label{eq:sm:action-current}
 \rhoA=\frac{A^*\partial_zA-A\partial_zA^*}{2\ii\mu},\qquad
 \jA=-\frac{\eps}{2\ii}(A^*\partial_tA-A\partial_tA^*),\qquad
 \partial_z\rhoA+\partial_t\jA=0,
\end{equation}
for either sign of $\eps$ and $\mu$. If $\jA\to0$ as $t\to\pm\infty$, then $\NA=\int_{\mathbb R}\rhoA\,\dd t$ is $z$ independent.

For the positive nondispersive medium of the main text, a plane wave $A=A_0\ee^{\ii(kz-\omega t)}$ in a time-independent medium with $\eps\omega^2=k^2/\mu$ carries
\begin{equation}
\label{eq:sm:monochromatic-action}
\jA=\frac{u}{\omega},\qquad \rhoA=\frac{S_z}{\omega},
\end{equation}
with $u=\tfrac12(\eps|E_x|^2+\mu|H_y|^2)$ the energy density and $S_z=\operatorname{Re}(E_xH_y^*)$ the longitudinal energy flux. Thus $\jA$ is the conventional wave-action density, the Noether charge density of the phase symmetry $A\mapsto\ee^{\ii\alpha}A$, and $\rhoA$ is its longitudinal flux~\cite{SMHayes1970,SMAndrewsMcIntyre1978}. Because $z$ is the evolution coordinate here, $\rhoA$ plays the role of a density along the fast time $t$.

\section{Comparison with waveguide pumps}
In a waveguide Thouless pump~\cite{SMKraus2012}, the paraxial envelope obeys a Schr\"odinger equation that is first order in \(z\). Its positive norm \(P=\int|\psi_{\rm wg}|^2\,\dd^2x_\perp\) is the optical power in the conventional normalization, and \(|\psi_{\rm wg}|^2\) is proportional to the longitudinal Poynting-flux density measured at the output. In the photonic time crystal (PTC) pump of this work, \(z\) remains the evolution coordinate and the periodic fast time \(t\) takes the transverse role. Table~\ref{tab:roles} compares the two pumps.

The second-order equation~\eqref{eq:sm:maxwell} requires the Cauchy pair \((A,\partial_zA)\). For a mode \(A=\ee^{\ii kz}\psi(t)\), the labels forward and backward refer to \(k>0\) and \(k<0\), respectively. The cell action equals
\begin{equation}
\label{eq:sm:branch-action}
 \int_0^T\rhoA\,\dd t=\frac{k}{\mu}\int_0^T|\psi|^2\,\dd t,
\end{equation}
so the branches carry opposite cell actions; we choose the positive-action branch. At a fixed positive carrier frequency, the forward-paraxial model gives proportional profiles with a common center [Eq.~\eqref{eq:sm:monochromatic-action}], although the quantities retain distinct dimensions and meanings. In the PTC, harmonics of both frequency signs contribute nonnegative cell action, while their signed-frequency weights cancel the energy flux for uniform full-zone occupation (Sec.~IV). Interference permits local sign changes in the Wannier density \(\rhoA(t)\).

\begin{table}[htb!]
\caption{Waveguide and PTC pump variables. The transverse current of the paraxial envelope is \(\mathbf j_\perp=\operatorname{Im}(\psi_{\rm wg}^*\nabla_\perp\psi_{\rm wg})/k_0\).}
\label{tab:roles}
\begin{ruledtabular}
\begin{tabular}{lll}
 & Waveguide pump & Pump of this work \\ \hline
Evolution coordinate & \(z\) & \(z\) \\
Transverse coordinate & \(\mathbf x_\perp\) & fast time \(t\); lattice period \(T\) \\
Order in \(z\) & first & second \\
Cauchy data & \(\psi_{\rm wg}\) & \((A,\partial_zA)\) \\
Conserved quantity & \(P=\int|\psi_{\rm wg}|^2\,\dd^2x_\perp\) & \(\NA=\int\rhoA\,\dd t\) \\
Transported center & \((1/P)\int \mathbf x_\perp|\psi_{\rm wg}|^2\,\dd^2x_\perp\) & \((1/\NA)\int t\rhoA\,\dd t\) \\
Transport rate & \(\dd\bar{\mathbf x}_\perp/\dd z=(1/P)\int\mathbf j_\perp\,\dd^2x_\perp\) & \(\dd\tA/\dd z=(1/\NA)\int\jA\,\dd t\) \\
Conserved-form signature & positive definite & indefinite; opposite on the \(\pm k\) branches \\
Adiabatic Wannier displacement & \(C\) spatial cells & \(\CB\) temporal cells \\
\end{tabular}
\end{ruledtabular}
\end{table}

\section{Positive-action band and the temporal Wannier center}
\begin{figure}[htb!]
\centering
\includegraphics[width=1\columnwidth]{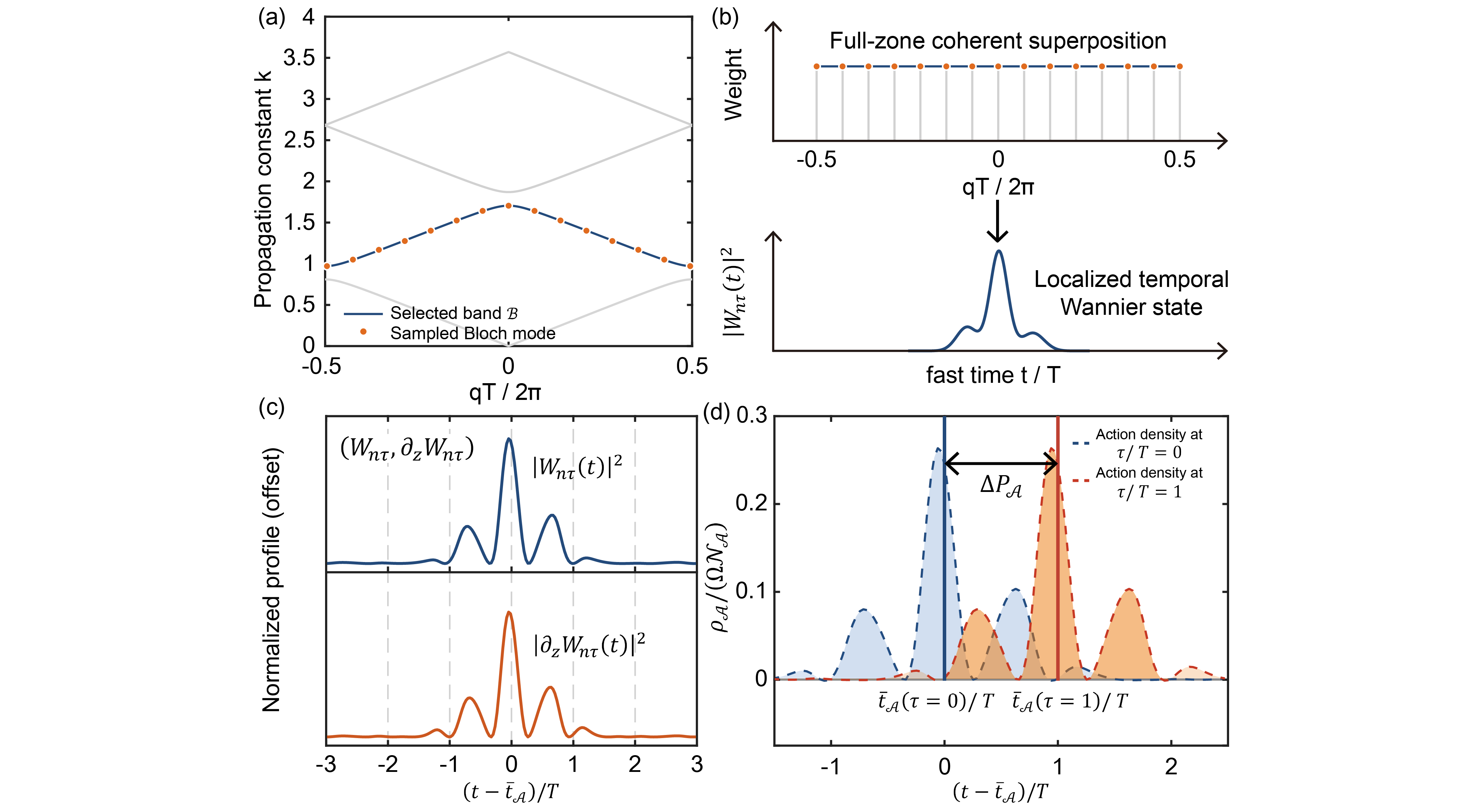}
\caption{\textbf{Construction of the temporal Wannier Cauchy state.}
(a) Instantaneous propagation bands over the temporal Bloch quasifrequency zone. The selected positive-action band $\mathcal B$ is highlighted in blue, and the markers indicate the quasifrequency sampling used for the full-zone construction.
(b) Schematic full-zone coherent superposition of the selected band and the resulting temporal Wannier profile localized in fast time.
(c) The two components of the Wannier Cauchy state, $W_{n\tau}$ and its branch-matched derivative $\partial_zW_{n\tau}$, shown as normalized squared magnitudes and vertically offset for clarity. Dashed lines indicate positions separated by one temporal period $T$.
(d) Instantaneous signed action-density profiles $\rho_{\mathcal A}/(\Omega\mathcal N_{\mathcal A})$ at $\tau/T=0$ and $1$ for the rigid cycle, shown in a common temporal reference frame. The vertical lines mark the corresponding action centers, separated by one temporal cell.}
\label{fig:s1}
\end{figure}

Let $\eps(t;\tau)>0$ be smooth and $T$ periodic in both $t$ and $\tau$, with $\mu>0$. Substituting $A=\ee^{\ii k_mz}\ee^{\ii qt}u_{mq\tau}(t)$, with $u_{mq\tau}(t+T)=u_{mq\tau}(t)$, into \eqref{eq:sm:maxwell} gives
\begin{equation}
\label{eq:sm:band-eigenproblem}
 \mathcal L_q[\tau]\,u_{mq\tau}=k_m^2\,u_{mq\tau},
 \qquad
 \mathcal L_q[\tau]\equiv-\mu\,D_q\,\eps(t;\tau)\,D_q,
 \qquad
 D_q\equiv\partial_t+\ii q.
\end{equation}
Integration by parts over one cell shows that $\mathcal L_q[\tau]=\mu D_q^\dagger\eps D_q$ is self-adjoint and, for $\eps>0$, nonnegative, so $k_m^2\ge0$. The eigenvalues form a ladder $k_m^2(q,\tau)$, and the projector onto an isolated simple band is smooth in $q$ and $\tau$. Select the positive-action branch of an isolated band $\mathcal B$ with a uniform gap and $k_{\mathcal B}\ge k_{\min}>0$. Its unit-action profile is
\begin{equation}
\label{eq:sm:unit-action-normalization}
 \int_0^T|u|^2\,\dd t=1,
 \qquad
 \widetilde\psi=(\mu/k_{\mathcal B})^{1/2}\,\ee^{\ii qt}u.
\end{equation}
With $\Omega=2\pi/T$, the exact identity
\begin{equation}
\label{eq:sm:zone-sewing}
 \mathcal L_{q+\Omega}[\tau]=U_\Omega\,\mathcal L_q[\tau]\,U_\Omega^\dagger,
 \qquad U_\Omega=\ee^{-\ii\Omega t},
\end{equation}
provides a periodic gauge with $u(q+\Omega)=U_\Omega u(q)$. At finite harmonic cutoff, the zone identification \eqref{eq:sm:zone-sewing} has a truncation error that decreases as the cutoff is increased. The Wannier profile and its branch-matched $z$ derivative form the Wannier Cauchy state [Fig.~\ref{fig:s1}(a)--(c)],
\begin{equation}
\label{eq:sm:wannier-cauchy-state}
 W_{n\tau}(t)=\frac{T}{2\pi}\int_{\BZ}\ee^{-\ii qnT}\widetilde\psi_{\mathcal Bq\tau}(t)\,\dd q,
 \qquad
 \partial_zW_{n\tau}(t)=\frac{T}{2\pi}\int_{\BZ}\ee^{-\ii qnT}\,\ii k_{\mathcal B}(q,\tau)\,\widetilde\psi_{\mathcal Bq\tau}(t)\,\dd q.
\end{equation}

Parallel transport, followed by a uniform distribution of the residual holonomy, gives a smooth $q$-periodic gauge. In this gauge, the restrictions of $W_{n\tau}$ to successive temporal cells are the Fourier coefficients of $\widetilde\psi_{\mathcal Bq\tau}$ with respect to $q$. Parseval's identity therefore gives
\begin{equation}
\label{eq:sm:parseval-action}
 \int_{\mathbb R}|W_{n\tau}|^2\,\dd t=\frac{T}{2\pi}\int_{\BZ}\frac{\mu}{k_{\mathcal B}}\,\dd q,
 \qquad
 \int_{\mathbb R}W_{n\tau}^*\,\partial_zW_{n\tau}\,\dd t=\ii\mu.
\end{equation}
The second identity gives $\NA[W_{n\tau}]=1$ exactly. With $t$ acting as $\ii\partial_q$ on the Fourier side, the same computation bounds $\|(t-nT)W_{n\tau}\|$ and $\|\partial_zW_{n\tau}\|$, so the first moment of $\rhoA$ converges absolutely.

At fixed $\tau$, write $k_q=k_{\mathcal B}(q,\tau)$ and suppress $\tau$ in the mode profiles. Substituting both components of \eqref{eq:sm:wannier-cauchy-state} into \eqref{eq:sm:action-current} and interchanging $q$ and $q'$ in the conjugate term combines the propagation constants and normalization factors into the symmetric coefficient
\begin{equation}
\label{eq:sm:action-kernel}
 \mathcal K(q,q')=\frac{k_q+k_{q'}}{2\sqrt{k_qk_{q'}}},
 \qquad
 \mathcal K(q,q)=1,
 \qquad
 \left.\partial_{q'}\mathcal K(q,q')\right|_{q'=q}=0.
\end{equation}
To evaluate the first moment, we use $t\ee^{\ii(q'-q)t}=-\ii\partial_{q'}\ee^{\ii(q'-q)t}$ and integrate by parts in $q'$. The boundary contribution cancels under the zone identification \eqref{eq:sm:zone-sewing}, and summation over temporal cells selects $q'=q$. The derivative of $\mathcal K$ then vanishes by \eqref{eq:sm:action-kernel}, the derivative of the Wannier phase supplies $nT$, and the derivative of the periodic factor gives $\ii\int_0^T u^*\partial_qu\,\dd t$. Together with the unit action established in \eqref{eq:sm:parseval-action}, this gives
\begin{equation}
\label{eq:sm:action-polarization}
 \tA=nT+\frac{T}{2\pi}\int_{\BZ}\BerryA_q\,\dd q,
 \qquad
 \BerryA_q=\ii\int_0^T u^*\partial_qu\,\dd t,
\end{equation}
with $\BerryA_q$ the Berry connection, real by normalization. Thus $\PA=(\tA-nT)/T$ is the Berry polarization modulo one cell. Figure~\ref{fig:s1}(d) illustrates the one-cell shift of the instantaneous center in a rigid cycle.

\begin{figure}[htb!]
\centering
\includegraphics[width=1.00\columnwidth]{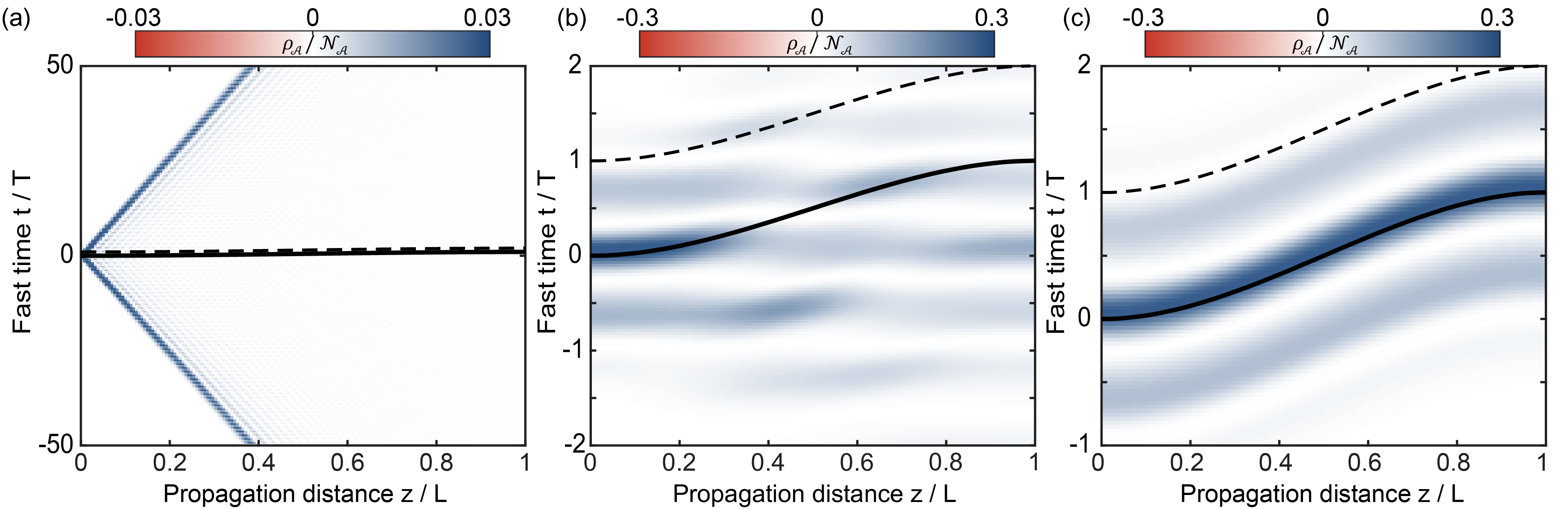}
\caption{\textbf{Three frames for the transported density.} Signed density \(\rho_{\mathcal A}(z,t)/\NA\) of the one-cell pump of Fig.~3 of the main text, propagated at \(L=480\,c/\Omega\) and shown in (a) the physical frame, (b) the propagation-phase-removed frame, and (c) the smooth rephased frame associated with \eqref{eq:sm:flat-connection}. The solid line is the action center, identical in the three panels; the dashed lines mark one temporal cell. In the physical frame, the different group delays of the quasifrequency components broaden the temporal distribution of wave action and reduce the density within each cell. The panels use different temporal ranges and color scales. The propagation protocol is specified in Sec.~VII.}
\label{fig:s2}
\end{figure}

The propagated field is reconstructed by coherent summation over quasifrequencies, with relative phases that determine the local action-density profile. The transformation
\begin{equation}
\label{eq:sm:cauchy-rephasing}
 \widetilde\psi_q(z)\;\longmapsto\;\ee^{\ii\vartheta(q,z)}\,\widetilde\psi_q(z),
\end{equation}
preserves the action of each component and \(\NA\), while changing the coherently reconstructed density \(\rhoA(z,t)\). The phase multiplies both entries of the propagated Cauchy pair, including its physical \(z\)-derivative component. For unit-action quasifrequency sectors, smooth \(q\)-periodic rephasings with zero winding also preserve the action center. Three frames are used in this work. The physical frame retains the original propagated phases, with \(\vartheta=0\). The propagation-phase-removed frame subtracts the accumulated band phase and, with it, the group delay,
\begin{equation}
\label{eq:sm:propagation-phase}
 \vartheta(q,z)=-\theta(q,z),
 \qquad
 \theta(q,z)=\int_0^z k_{\mathcal B}[q,\tau(z')]\,\dd z',
\end{equation}
for a device that traverses the cycle as \(\tau(z)\). The third frame applies a smooth \(q\)-dependent rephasing to the complete propagated Cauchy pair, with the residual holonomy distributed uniformly over the zone. For the instantaneous band, the corresponding phase convention makes the connection independent of \(q\),
\begin{equation}
\label{eq:sm:flat-connection}
 \BerryA_q=\frac{T}{2\pi}\int_{\BZ}\BerryA_{q'}\,\dd q',
\end{equation}
giving a smooth periodic band representation. Equation~\eqref{eq:sm:flat-connection} concerns the instantaneous-band factor \(u\); the frame transformation of the propagated state acts on its complete Cauchy pair. The main-text maps use this smooth rephased representation. Figure~\ref{fig:s2} compares density profiles of the same one-cell pump in the three frames with a common action center.

\section{Vanishing of the energy flux and its first moment}
A single harmonic obeys \(S_z=\omega\rhoA\), as in \eqref{eq:sm:monochromatic-action}. The field-energy balance \(\partial_tu+\partial_zS_z=-\tfrac12(\partial_t\eps)|E_x|^2\), integrated over the fast time, gives \(\dd/\dd z\int S_z\,\dd t=-\tfrac12\int(\partial_t\eps)|E_x|^2\,\dd t\). We expand the periodic factor as \(u_{\mathcal Bq}=\sum_\ell a_\ell(q)\ee^{\ii\ell\Omega t}\), where harmonic \(\ell\) has physical frequency \(\omega_\ell=-(q+\ell\Omega)\). Integration over one temporal cell removes the cross terms and gives the harmonic contributions to the action and energy flux of the unit-action state:
\begin{equation}
\label{eq:sm:harmonic-contributions}
 \int_0^T\rhoA\,\dd t=T\sum_\ell|a_\ell|^2=1,
 \qquad
 \int_0^T S_z\,\dd t=T\sum_\ell\omega_\ell|a_\ell|^2.
\end{equation}

\begin{figure}[htb!]
\centering
\includegraphics[width=1\columnwidth]{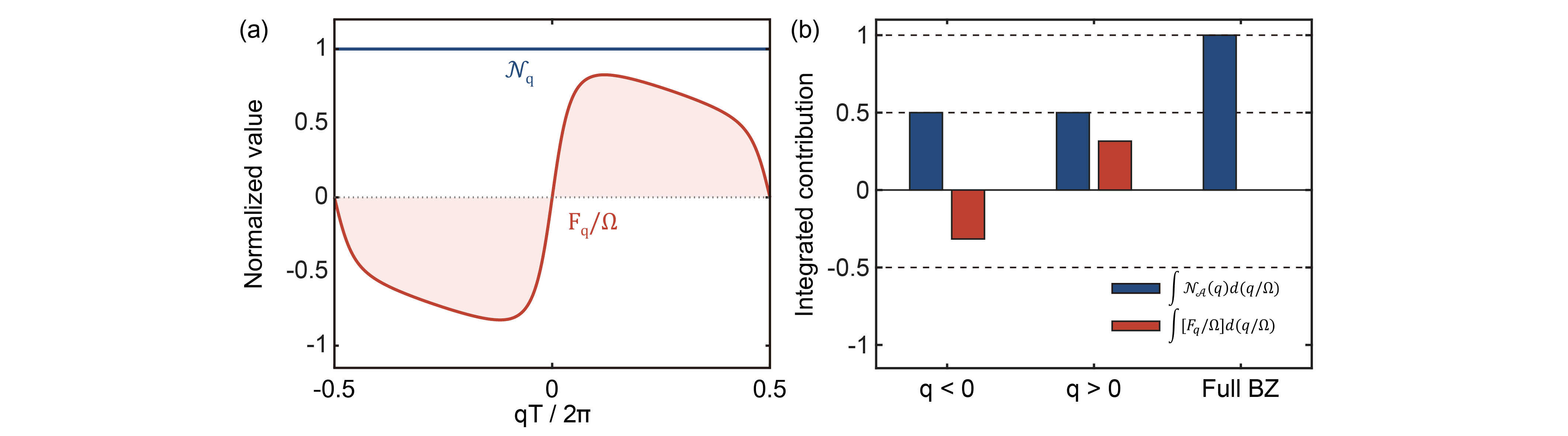}
\caption{\textbf{Full-zone parity and energy-flux cancellation.}
(a) Quasifrequency-resolved cell action $\mathcal N_{\mathcal A}(q)$
and action-normalized energy flux $F_q/\Omega$ for the selected
positive-action band. The cell action is even in $q$, whereas the
energy flux is odd.
(b) Corresponding integrated contributions from the negative and
positive halves of the quasifrequency zone and from the full zone.
The two half-zone action contributions add to unity, whereas the
equal-and-opposite energy-flux contributions cancel over the full
zone.}
\label{fig:s3}
\end{figure}

The \(z\) derivative contributes the same propagation constant \(k_{\mathcal B}\) to every harmonic of a Bloch mode, whereas the \(t\) derivative contributes the corresponding harmonic frequency. The ratio of the energy-flux integral to the action integral in \eqref{eq:sm:harmonic-contributions} is therefore the action-weighted mean physical frequency:
\begin{equation}
\label{eq:sm:flux-per-action}
 F_q\equiv
 \frac{\int_0^T S_z\,\dd t}{\int_0^T\rhoA\,\dd t}
 =-\frac{\sum_\ell(q+\ell\Omega)|a_\ell|^2}
 {\sum_\ell|a_\ell|^2}.
\end{equation}
On the selected \(+k\) branch, all diagonal harmonic contributions to the cell action are nonnegative, consistent with the positive Krein signature of the band. Each energy-flux contribution is additionally weighted by its signed physical frequency. The same signed weighting appears in the photon-number orthogonality relation for a time-modulated grating~\cite{SMPendry2023}. Reality of \(\eps\) and \(\mu\) relates the \((q,\ell)\) and \((-q,-\ell)\) components through \(u_{\mathcal B,-q}=\ee^{\ii\chi_q}u_{\mathcal Bq}^*\), so the action is even in \(q\) and the energy flux is odd. Figure~\ref{fig:s3}(a) shows these parities; panel (b) separates the negative- and positive-half-zone integrals. Uniform occupation of the full zone therefore gives
\begin{equation}
\label{eq:sm:zero-net-flux}
 \int_{\mathbb R}S_z\,\dd t
 =\frac{T}{2\pi}\int_{\BZ}F_q\,\dd q
 =0.
\end{equation}
The first moment follows from the same reduction as \eqref{eq:sm:action-polarization}:
\begin{equation}
\label{eq:sm:flux-first-moment}
 \int_{\mathbb R}tS_z\,\dd t
 =\frac{T}{2\pi}\int_{\BZ}\!\left[nT\,F_q-G_q\right]\dd q,
 \qquad
 G_q=\operatorname{Im}\int_0^T
 u^*(\hat\omega-q)\partial_qu\,\dd t,
 \qquad \hat\omega=\ii\partial_t.
\end{equation}
The first term vanishes by \eqref{eq:sm:zero-net-flux}. The relation \(G_{-q}=-G_q+\chi_q'F_q\) makes \(G_q\) odd when \(\chi_q\) is independent of \(q\); the first moment therefore vanishes in that frame. Since the zeroth moment in \eqref{eq:sm:zero-net-flux} vanishes in every gauge, the normalized full-zone energy-flux center is undefined. For partial-zone occupation with nonzero integrated flux, the normalized flux center depends on the chosen occupation. Finite propagation leaves a small residual that decreases approximately as \(L^{-1}\) over the range shown in Fig.~2(d) of the main text.

The positive electromagnetic-energy density also admits a normalized center \(\bar t_E\). A unit-action Bloch mode carries the cell energy
\begin{equation}
\label{eq:sm:mode-energy}
 \mathcal E_q=\frac12\int_0^T\Big[\eps|\partial_t\widetilde\psi|^2+\frac{k_{\mathcal B}^2}{\mu}|\widetilde\psi|^2\Big]\dd t
 =\frac{k_{\mathcal B}}2+\frac{k_{\mathcal B}}2=k_{\mathcal B}(q),
\end{equation}
The eigenvalue equation gives equal electric and magnetic contributions, each \(k_{\mathcal B}/2\). The energy center is weighted by the positive band energies \(k_{\mathcal B}(q)\). Under a periodic \(q\)-rephasing \(u_q\mapsto\ee^{\ii\varphi(q)}u_q\) with no winding across the zone,
\begin{equation}
\label{eq:sm:rephased-centers}
 \delta\tA=-\frac{T}{2\pi}\int_{\BZ}\partial_q\varphi\,\dd q=0,
 \qquad
 \delta\bar t_E=-\frac{\int_{\BZ}k_{\mathcal B}\,\partial_q\varphi\,\dd q}{\int_{\BZ}k_{\mathcal B}\,\dd q},
\end{equation}
so the action center is fixed, whereas the energy center generally changes. The corresponding first-moment reduction weights the Berry connection by \(k_{\mathcal B}\); hence the \(q\)-independent connection in \eqref{eq:sm:flat-connection} gives coincident energy and action centers.

\section{Chern transport and the rigid one-cell pump}
Over a closed cycle, the change in the polarization of Sec.~III is quantized. On the pump torus \((q,\tau)\in S_q^1\times S_\tau^1\), define
\begin{equation}
\label{eq:sm:berry-curvature}
 \BerryF_{\tau q}
 =\partial_\tau\BerryA_q-\partial_q\BerryA_\tau,
 \qquad
 \BerryA_\tau=\ii\int_0^T u^*\partial_\tau u\,\dd t .
\end{equation}
In a \(q\)-periodic gauge, the polarization derivative is \(\dd\PA/\dd\tau=(1/2\pi)\int_{\BZ}\BerryF_{\tau q}\,\dd q\). Integrating over one cycle with orientation \(\dd\tau\wedge\dd q\) gives
\begin{equation}
\label{eq:sm:chern-transport}
 \frac{\tA(T)-\tA(0)}{T}
 =\CB
 =\frac{1}{2\pi}\int_0^T\!\dd\tau\int_{\BZ}\!\dd q\,\BerryF_{\tau q}
 =\frac{1}{2\pi\ii}\int
 \Tr\!\left\{P_{\mathcal B}
 [\partial_qP_{\mathcal B},\partial_\tau P_{\mathcal B}]
 \right\}\dd\tau\dd q
 \in\mathbb Z .
\end{equation}
Here \(\CB\) is the Chern number of the instantaneous band~\cite{SMThouless1983}, and \(P_{\mathcal B}=|u\rangle\langle u|\) is its projector. A nonzero \(\CB\) precludes a single smooth periodic frame on the torus, so the band is covered by local gauges related by transition phases. The smooth embedding of the Bloch profile into its Cauchy pair is invertible on the isolated band, so the Cauchy line bundle has the same Chern number.

For a rigid translation \(\eps_{\rm r}(t;\tau)=\eps_0(t-\tau)\), choose the Bloch functions as \(u_{\mathcal Bq\tau}(t)=\ee^{-\ii q\tau}u_{\mathcal Bq0}(t-\tau)\). Substitution into the Berry connection in \eqref{eq:sm:action-polarization} gives
\begin{equation}
\label{eq:sm:rigid-pump}
 \BerryA_q(q,\tau)=\BerryA_q(q,0)+\tau,
 \qquad
 \PA(\tau)=\PA(0)+\tau/T,
 \qquad
 \CB=+1.
\end{equation}
The same result follows from the cycle-end transition function \(\chi(q)=\ee^{-\ii qT}\), whose winding is \(-1\). Equation~\eqref{eq:sm:rigid-pump} gives the instantaneous Wannier center; the propagated center differs by the finite-length \(O(L^{-1})\) correction in \eqref{eq:sm:finite-length-center}. On a discrete \(q\) grid,
\begin{equation}
\label{eq:sm:wilson-loop}
 \PA(\tau)
 =-\frac{1}{2\pi}\operatorname{Im}
 \ln\prod_j
 \langle u(q_j,\tau)|u(q_{j+1},\tau)\rangle .
\end{equation}
The terminal overlap includes $U_\Omega$ to implement the physical zone identification $u(q_1+\Omega)=U_\Omega u(q_1)$ of \eqref{eq:sm:zone-sewing}~\cite{SMFukui2005}.

Figures 2(b)--(d) and 3 of the main text use the same rigid-pump medium with $\eps_b=3.4$, $\hat\eps_1(0)=0.70$, and $\hat\eps_2(0)=0.30\,\ee^{-0.30\ii}$, with the second forward band as the target. Figures 2(b) and 2(c) are evaluated at $\tau=0$, whereas Fig. 2(d) and Fig. 3 traverse the rigid cycle.

\section{Intrinsic single-band \texorpdfstring{$C_{\mathcal B}=2$}{CB=2} cycle}
The two-tone cycle varies the two harmonics of $\eps$ independently and realizes $\CB=2$ in a single band. Figure~4 of the main text uses
\begin{equation}
\label{eq:sm:two-tone-parameters}
 (\eps_b,g,a,R_s,R_r)
 =\left(\tfrac92,\tfrac32,\tfrac1{10},\tfrac34,\tfrac12\right),
 \qquad
 s=R_s\cos\Omega\tau,
 \qquad
 r=-R_r\sin\Omega\tau .
\end{equation}
It is convenient to write
\begin{equation}
\label{eq:sm:two-tone-medium}
\begin{aligned}
 \bar\eps(t;s,r)&\equiv\eps/\eps_b=1+\eta f(t;s,r),
 \qquad \eta=g/\eps_b=\tfrac13,\\
 f&=(s-a)\cos\Omega t+r\sin\Omega t
 +(s+a)\cos2\Omega t-r\sin2\Omega t .
\end{aligned}
\end{equation}
On the closed control disk, \(|f|\le3/2\), so \(\eps\ge9/4>0\). The possible quasifrequencies of a band degeneracy follow from the scalar second-order structure of \eqref{eq:sm:band-eigenproblem}. At fixed control parameters $(s,r)$, the Bloch solution $\psi=\ee^{\ii qt}u$ obeys $\partial_t(\eps\partial_t\psi)+(k^2/\mu)\psi=0$. Its Cauchy vector $\mathbf y=(\psi,\eps\partial_t\psi)^{\mathsf T}$ therefore satisfies
\begin{equation}
\label{eq:sm:cauchy-generator}
 \partial_t\mathbf y
 =
 \begin{pmatrix}
  0 & \eps^{-1}(t;s,r)\\
  -k^2/\mu & 0
 \end{pmatrix}
 \mathbf y .
\end{equation}
The one-period monodromy $\mathsf M(k^2)$ is defined by $\mathbf y(T)=\mathsf M(k^2)\mathbf y(0)$. The coefficient matrix in~\eqref{eq:sm:cauchy-generator} is traceless, so Liouville's formula gives $\det\mathsf M(k^2)=1$.

For a Bloch solution, periodicity of both $u$ and $\eps$ implies $\mathbf y(T)=\ee^{\ii qT}\mathbf y(0)$. Thus its initial Cauchy vector is an eigenvector of $\mathsf M(k^2)$ with multiplier $\ee^{\ii qT}$. At a degenerate eigenvalue of the self-adjoint operator in \eqref{eq:sm:band-eigenproblem}, there are two linearly independent eigenfunctions. The corresponding Bloch solutions have linearly independent initial Cauchy vectors, since dependent initial data would generate dependent solutions by uniqueness of the Cauchy problem. These vectors span the two-dimensional Cauchy space and have the same Bloch multiplier, so the monodromy at the degenerate eigenvalue is $\ee^{\ii qT}\openone$. Together with its unit determinant, this gives
\begin{equation}
\label{eq:sm:degeneracy-quasifrequencies}
 \mathsf M=\ee^{\ii qT}\openone,
 \qquad
 \det\mathsf M=\ee^{2\ii qT}=1,
 \qquad
 q\in\{0,\,\Omega/2\}\quad(\mathrm{mod}\ \Omega).
\end{equation}
The search for band degeneracies therefore reduces to the periodic and antiperiodic Bloch sectors. The two harmonics of \eqref{eq:sm:two-tone-medium} carry the amplitudes
\begin{equation}
\label{eq:sm:harmonic-amplitudes}
 \alpha\equiv\frac{\hat\eps_1}{g}=\tfrac12[(s-a)-\ii r],
 \qquad
 \beta\equiv\frac{\hat\eps_2}{g}=\tfrac12[(s+a)+\ii r].
\end{equation} At \(\alpha=0\) the first harmonic vanishes and the medium is exactly \(T/2\) periodic, so the half-period translation \(\mathsf S\) commutes with \eqref{eq:sm:band-eigenproblem}. In the \(q=\Omega/2\) sector,
\begin{equation}
\label{eq:sm:half-period-translation}
 \mathsf S\psi(t)=\psi(t+T/2),
 \qquad
 \mathsf S^2=\ee^{\ii qT}=-1,
\end{equation}
so the eigenvalues of \(\mathsf S\) are \(\pm\ii\). Complex conjugation maps the eigenspace of \(\mathsf S\) with eigenvalue \(+\ii\) to the eigenspace with eigenvalue \(-\ii\), while preserving the eigenvalue of \eqref{eq:sm:band-eigenproblem}. Each level is therefore doubly degenerate at \(\mathsf N_1\). The first harmonic couples the two symmetry sectors away from \(\alpha=0\).

\begin{figure}[htb!]
\centering
\includegraphics[width=1\columnwidth]{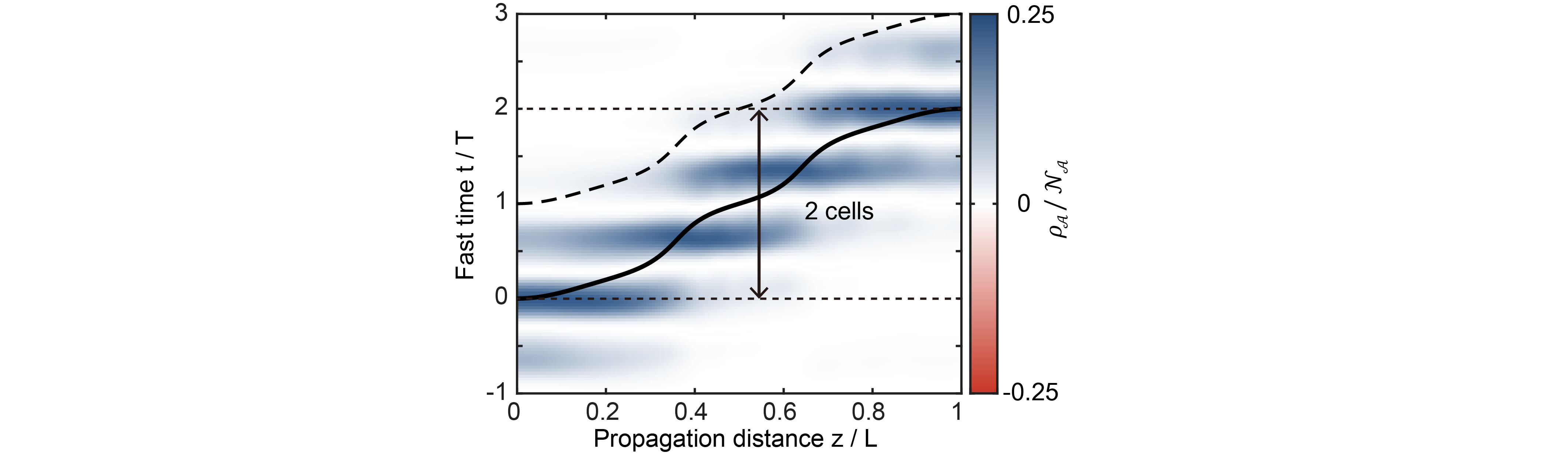}
\caption{\textbf{Two-cell action transport.} Signed density \(\rho_{\mathcal A}(z,t)/\NA\) for the two-tone Wannier Cauchy state propagated to \(L=960\,c/\Omega\), with the quasifrequency-dependent propagation phase removed. The action center advances by \(2.003\) cells; the raw physical-frame center gives the same displacement. The magnitude of the negative part of the plotted action density remains below \(0.5\%\) of its peak value.}
\label{fig:s4}
\end{figure}

\begin{figure}[htb!]
\centering
\includegraphics[width=1\columnwidth]{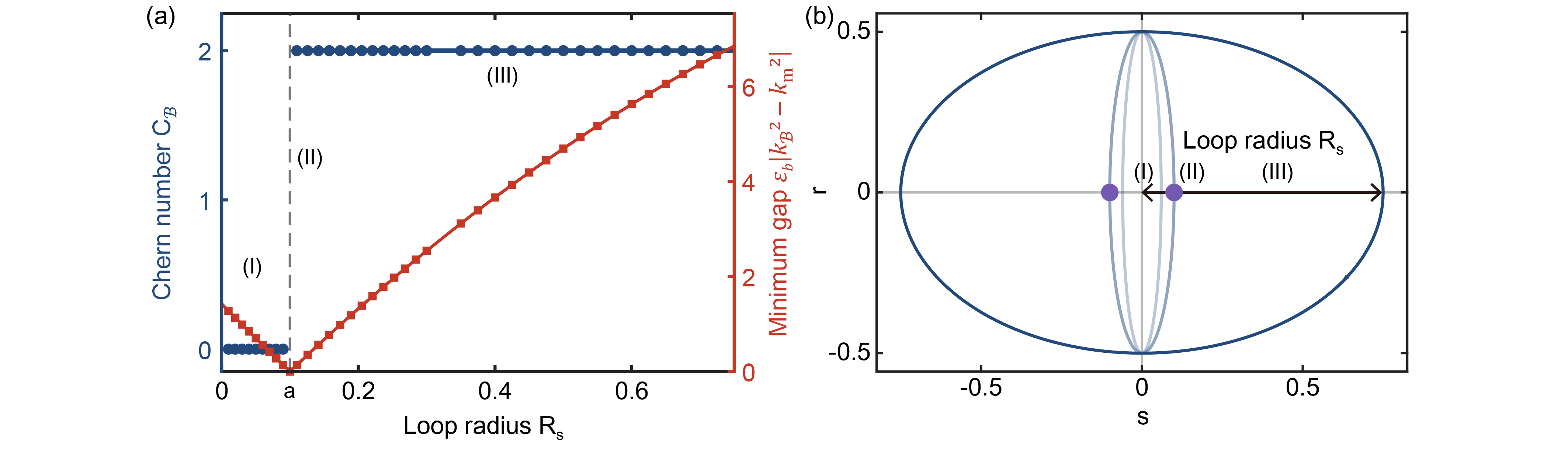}
\caption{\textbf{Chern plateau.} (a) Target-band Chern number and minimum squared-propagation-constant gap over the cycle versus \(R_s\), at \(R_r=1/2\) and \(a=1/10\). Within the displayed range, the gap closes at \(R_s=a\), separating the \(\CB=0\) and \(\CB=2\) regimes. (b) The projections of the two symmetry-enforced nodes onto the $(s,r)$ plane lie at $(\pm a,0)$. The loop encloses neither node for \(R_s<a\); it encloses both for \(R_s>a\).}
\label{fig:s5}
\end{figure}

To resolve the degeneracy at \(\mathsf N_2\), use the normalized Fourier basis \(e_\ell(t)=T^{-1/2}\ee^{\ii\ell\Omega t}\) and write \(\eps(t;s,r)=\sum_j\hat\eps_j\ee^{\ii j\Omega t}\). At \(q=0\), the operator in~\eqref{eq:sm:band-eigenproblem} has matrix elements
\begin{equation}
\label{eq:sm:n2-fourier-matrix}
 \left\langle e_\ell,\mathcal L_0 e_{\ell'}\right\rangle
 =\mu\,\ell\ell'\Omega^2\,\hat\eps_{\ell-\ell'}.
\end{equation}
At \(\beta=0\), only the first harmonic remains, so off-diagonal entries connect adjacent indices. The factor \(\ell\ell'\) sets the zero-index row and column to zero, separating the positive and negative harmonic sectors from the constant \(k^2=0\) mode. For real \(\eps\), complex conjugation maps either nonzero sector onto the other with the same eigenvalue. The conjugate eigenfunctions occupy disjoint sectors and are independent, giving the paired levels at \(\mathsf N_2\). The two nodes are located at
\begin{equation}
\label{eq:sm:node-locations}
 \mathsf N_1=(\Omega/2,\,a,\,0),
 \qquad
 \mathsf N_2=(0,\,-a,\,0).
\end{equation}
The target-pair splitting is positive over the closed control disk \(\overline{\mathcal D}=\{(s,r):\,s^2/R_s^2+r^2/R_r^2\le 1\}\) at both special quasifrequencies, outside two balls centered on \(\mathsf N_1\) and \(\mathsf N_2\). The target line remains isolated on the pump torus and on the two enclosing spheres. We define the local degree as the first Chern number of the lower eigenline on a small outward-oriented sphere surrounding each node. The projector onto the crossing pair extends smoothly over a sufficiently small enclosing ball, where the pair remains separated from the other bands. Its total first Chern number on the boundary sphere is zero, so the upper and lower eigenlines carry opposite local degrees. On outward-oriented spheres in the \((q,s,r)\) coordinates, the lower pair member has local degree \(-1\) at \(\mathsf N_1\) and \(+1\) at \(\mathsf N_2\). At \(\mathsf N_1\), the target line is the upper member and therefore contributes \(+1\). At \(\mathsf N_2\), it is the lower member and also contributes \(+1\). Adding the two target-band contributions gives
\begin{equation}
\label{eq:sm:two-cell-chern}
 \CB=1+1=2.
\end{equation}
In Fig.~\ref{fig:s4}, finite-length propagation shifts the action center by approximately two temporal cells, consistent with $\CB=2$. A full-torus plaquette sum yields the same integer. Over the range shown in Fig.~\ref{fig:s5}, scanning \(R_s\) at fixed \(R_r\) gives \(\CB=0\) for \(R_s<a\), a gap closing at \(R_s=a\), and \(\CB=2\) for \(R_s>a\).

\section{Finite-length transport and leakage}
\begin{figure}[htb!]
\centering
\includegraphics[width=1\columnwidth]{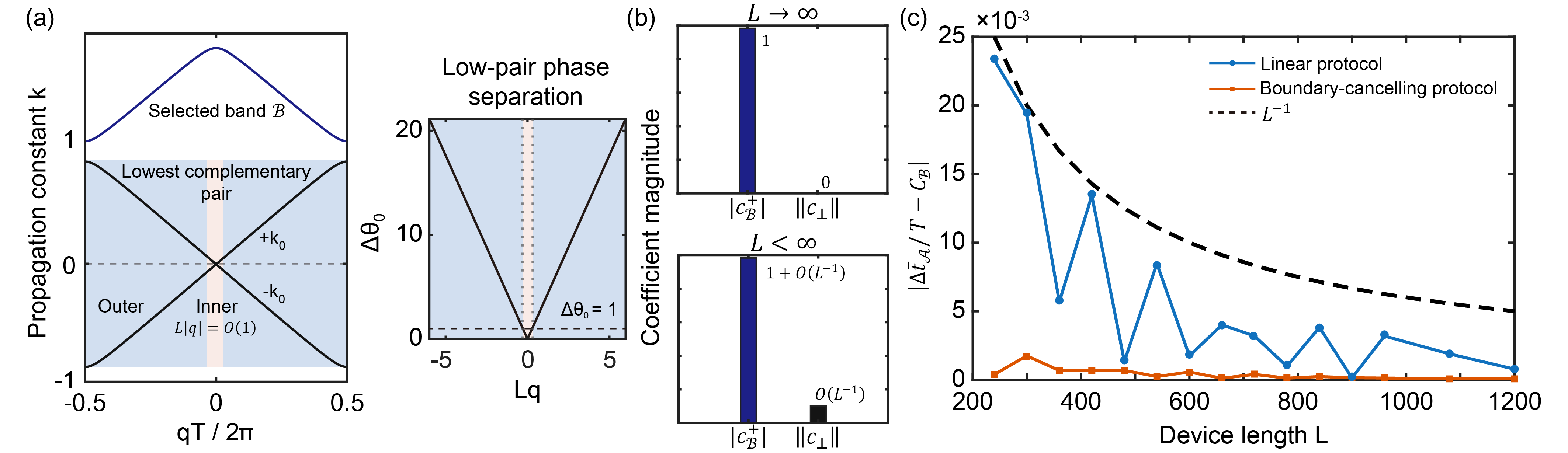}
\caption{\textbf{Finite-length adiabatic structure and convergence.}
(a) Schematic propagation spectrum near the lowest complementary pair. 
The pair $\pm k_0$ merges at $q=0$; the inner region $L|q|=O(1)$ is treated as a smooth Cauchy block, while the outer region is described by the signed-branch estimate. The right panel shows the relative phase $\Delta\theta_0=\theta_0^+-\theta_0^-$ versus $Lq$; the line $\Delta\theta_0=1$ indicates the crossover scale.
(b) Modal amplitudes in the adiabatic limit and at finite $L$. As $L\to\infty$, $|c_{\mathcal B}^{+}|\to1$ and $\|\bm c_\perp\|\to0$; at finite $L$, $|c_{\mathcal B}^{+}|=1+O(L^{-1})$ and $\|\bm c_\perp\|=O(L^{-1})$.
(c) Finite-length error $|\Delta\bar t_{\mathcal A}/T-C_{\mathcal B}|$ for the linear and boundary-cancelling protocols. The dashed reference curve is proportional to $L^{-1}$; boundary cancellation suppresses the finite-length error over the displayed range.}
\label{fig:s6}
\end{figure}

The pump varies with \(z\), coupling instantaneous branches, while its \(T\)-periodicity keeps the quasifrequency sectors independent. Write \(\zeta=z/L\) and traverse one cycle as
\begin{equation}
\label{eq:sm:pump-schedule}
\tau(\zeta)=T\,h(\zeta),\qquad h(0)=0,\qquad h(1)=1,\qquad h\in C^3([0,1]).
\end{equation}
We work at a fixed harmonic cutoff. For a symmetric harmonic window the values of \(k_{\mathcal B}\) at the zone endpoints \(q=\pm\Omega/2\) coincide exactly, since reality of \(\eps\) makes \(k_{\mathcal B}\) even in \(q\). The identification \eqref{eq:sm:zone-sewing} then holds up to a tail that decays with the cutoff. The dynamical center estimate below uses sewn Bloch data; the finite-window identification error is a separate reconstruction error. For the regular complement, we assume uniform bounds on the selected-to-complement couplings, branch modes, cell-center matrix elements, and inverse signed gaps. These bounds hold throughout \(\BZ\times[0,1]\) and also apply to their \(q\) and \(\zeta\) derivatives. The selected band remains separated from \(k=0\), ensuring regular unit-action normalization in \eqref{eq:sm:unit-action-normalization}. The propagation constants of the lowest complementary pair vanish as \(q\to0\), where the two signed branches merge. We represent this pair in a smooth two-dimensional Cauchy block and analyze it at the end of the section. Figure~\ref{fig:s6}(a) summarizes this spectral structure and the inner--outer decomposition used below.

The exact Cauchy state at each \(q\) is expanded over the instantaneous signed branch modes in the cell action pairing \(\kpair{X}{Y}=\int_0^T(X_1^*Y_2-X_2^*Y_1)/(2\ii\mu)\,\dd t\). For a regular branch label \(i=(m,\sigma_i)\), let \(\lambda_i=\sigma_i k_m\) and choose \(\kpair{v_i}{v_j}=\sigma_i\delta_{ij}\). After extracting the dynamical phases,
\begin{equation}
\label{eq:sm:coupled-amplitudes}
X_q(\zeta)=\sum_i c_i(q,\zeta)\,v_i(q,\zeta)\,\ee^{\ii\theta_i(q,\zeta)},
\qquad
\theta_i(q,\zeta)=L\int_0^\zeta\lambda_i(q,s)\,\dd s,
\qquad
\partial_\zeta c_i
=-\sigma_i\sum_j
\kpair{v_i}{\partial_\zeta v_j}\,
\ee^{\ii(\theta_j-\theta_i)}c_j .
\end{equation}
We use \(\zeta\)-parallel transport, so the diagonal connection vanishes. The initial Wannier state occupies only the selected forward branch. For the lowest complementary pair near \(q=0\), Eq.~\eqref{eq:sm:coupled-amplitudes} is replaced by the smooth two-dimensional Cauchy-block equation described below. Each relative phase accumulates at the rate set by the corresponding signed propagation gap. Define the local gap and its cycle minimum by
\begin{equation}
\label{eq:sm:signed-gap}
g_k(q,\tau)=\min_{(m,\sigma)\ne(\mathcal B,+)}\big|k_{\mathcal B}-\sigma k_m\big|,
\qquad
g_*=\min_{q,\tau}g_k(q,\tau)>0.
\end{equation}
\begin{samepage}
For a transition from the selected band to a regular complementary branch, write \(\theta_{\mathcal B}-\theta_i=L\Phi\), so that \(\Phi'=k_{\mathcal B}-\lambda_i\) and \(|\Phi'|\ge g_*\). For a smooth prefactor \(F\), integration by parts gives
\begin{equation}
\label{eq:sm:oscillatory-ibp}
\begin{aligned}
\int_0^\zeta F(s)\,\ee^{\ii L\Phi(s)}\,\dd s
={}&\left[\frac{F(s)\,\ee^{\ii L\Phi(s)}}{\ii L\Phi'(s)}\right]_0^\zeta\\
&-\frac{1}{\ii L}\int_0^\zeta
\ee^{\ii L\Phi(s)}\,
\frac{\dd}{\dd s}\!\left(\frac{F(s)}{\Phi'(s)}\right)\dd s .
\end{aligned}
\end{equation}
\end{samepage}
The boundary term and the remaining integral are \(O(L^{-1})\) when \(F/\Phi'\) and its derivative have bounds independent of \(L\). In the coupled amplitude equations, the integral inequalities below control the mode amplitudes entering the prefactor \(F\).

The exit leakage is the unsigned modal action weight outside the selected forward branch,
\begin{equation}
\label{eq:sm:off-band-weight}
W_\perp(L)=\frac{T}{2\pi}\int_{\BZ}\Big[\sum_{m\ne\mathcal B}|c_m^{+}(q,1)|^2+\sum_m|c_m^{-}(q,1)|^2\Big]\dd q .
\end{equation}
Applying \eqref{eq:sm:oscillatory-ibp} to the gap-controlled terms yields coupled integral inequalities with \(O(L^{-1})\) source terms. Iterating these inequalities gives an exponential factor (Gr\"onwall's inequality) bounded independently of \(L\) on \(0\le\zeta\le1\). Hence, uniformly in \(q\) and \(\zeta\),
\begin{equation}
\label{eq:sm:generic-amplitudes}
\bm c_\perp=O(L^{-1}),
\qquad
c_{\mathcal B}^{+}=1+O(L^{-1}),
\qquad
W_\perp(L)=O(L^{-2}).
\end{equation}
The corresponding finite-length modal decomposition is illustrated schematically in Fig.~\ref{fig:s6}(b).

For the moment calculation, let \(X_q=(a_q,b_q)^{\mathsf T}\) denote the periodic Cauchy factors, with the common \(\ee^{\ii qt}\) removed and the physical \(z\) derivative retained as the second component. The selected unit-action vector is \(v_{\mathcal B}=(\mu/k_{\mathcal B})^{1/2}(u_{\mathcal B},\ii k_{\mathcal B}u_{\mathcal B})^{\mathsf T}\), with \(\kpair{v_{\mathcal B}}{\partial_\zeta v_{\mathcal B}}=0\) and \(\ii\kpair{v_{\mathcal B}}{\partial_qv_{\mathcal B}}=\BerryA_q\). Remove the common selected-band phase from both components and write
\begin{equation}
\label{eq:sm:state-decomposition}
Y_q=\ee^{-\ii\theta(q,L\zeta)}X_q=c\,v_{\mathcal B}+r,
\qquad c=c_{\mathcal B}^{+},\qquad
\kpair{v_{\mathcal B}}{r}=0,
\qquad |c|^2-1=-\kpair{r}{r}.
\end{equation}
The last identity in~\eqref{eq:sm:state-decomposition} follows from exact unit action in each \(q\) sector. We measure remainders in the positive cell norm
\begin{equation}
\label{eq:sm:cauchy-norm}
\|X\|_{\mathrm C}^2=\int_0^T\bigl(|X_1|^2+\ell_0^2|X_2|^2\bigr)\,\dd t,
\qquad \ell_0>0\ \text{fixed}.
\end{equation}
Differentiation of an oscillatory phase with respect to \(q\) introduces a factor proportional to \(L\). In the complementary mixing terms, this factor multiplies an amplitude of order \(L^{-1}\), giving a bounded contribution. For the direct selected-band source, integration by parts using the gap in \eqref{eq:sm:signed-gap} gives a bounded term. Applying the same integral-inequality argument to the differentiated system, with remainders measured in the norm~\eqref{eq:sm:cauchy-norm}, gives
\begin{equation}
\label{eq:sm:moment-derivative-bounds}
c-1=O(L^{-1}),\qquad \|r\|_{\mathrm C}=O(L^{-1}),\qquad
\partial_qc=O(1),\qquad \|\partial_qr\|_{\mathrm C}=O(1).
\end{equation}
The low-pair rescaling below makes the bounds in~\eqref{eq:sm:moment-derivative-bounds} uniform through \(q=0\).

Exact unit action in each \(q\) sector reduces the common-phase contribution to \(-(T/2\pi)\int_{\BZ}\partial_q\theta\,\dd q\). This integral vanishes because \(\theta\) is periodic across the zone. Differentiating the action orthogonality in~\eqref{eq:sm:state-decomposition} gives \(\kpair{v_{\mathcal B}}{\partial_qr}=-\kpair{\partial_qv_{\mathcal B}}{r}\), so the Bloch--Parseval reduction of \eqref{eq:sm:action-polarization} yields
\begin{equation}
\label{eq:sm:moment-expansion}
\begin{aligned}
\frac{\tA-\tA^{\rm inst}}{T}
=\frac{1}{2\pi}\int_{\BZ}\!\dd q\,\Bigl\{
&(|c|^2-1)\BerryA_q+\operatorname{Re}(\ii c^*\partial_qc)\\
&+2\operatorname{Im}\!\left[c^*\kpair{\partial_qv_{\mathcal B}}{r}\right]
+\operatorname{Re}\!\left[\ii\kpair{r}{\partial_qr}\right]\Bigr\}.
\end{aligned}
\end{equation}
Here \(\tA^{\rm inst}=nT+(T/2\pi)\int_{\BZ}\BerryA_q\,\dd q\). The first term in~\eqref{eq:sm:moment-expansion} is \(O(L^{-2})\). Since \(c\) is periodic in the sewn gauge,
\(\int_{\BZ}c^*\partial_qc\,\dd q=\int_{\BZ}(c^*-1)\partial_qc\,\dd q=O(L^{-1})\).
The term linear in \(r\) is \(O(L^{-1})\), since \(\partial_qv_{\mathcal B}\) is bounded and \(r=O(L^{-1})\). The quadratic term has the same order because it combines \(r=O(L^{-1})\) with \(\partial_qr=O(1)\). Thus, for a generic \(C^3\) schedule,
\begin{equation}
\label{eq:sm:finite-length-center}
\frac{\tA(\zeta)-\tA^{\rm inst}[\tau(\zeta)]}{T}=O(L^{-1}),
\qquad
\frac{\tA(1)-\tA(0)}{T}=\CB+O(L^{-1}),
\end{equation}
where the second equality follows on the continuously unwrapped branch of \eqref{eq:sm:chern-transport}. For a numerically reconstructed center, \eqref{eq:sm:finite-length-center} acquires an error controlled by the absolute zeroth and first moments of the density error, including the omitted temporal tails. If the reconstructed action stays bounded away from zero and these moments tend to zero at the input and output, the reconstructed displacement converges.

The center can be written as the sum of dynamical and geometric contributions:
\begin{equation}
\label{eq:sm:center-decomposition}
\frac{\tA(\zeta)}{T}=n+\frac{1}{2\pi}\int_{\BZ}\big[-\partial_q\theta(q,\zeta L)+\BerryA_q(q,\tau(\zeta))\big]\,\dd q+O(L^{-1}).
\end{equation}
Here \(\theta\) is given by \eqref{eq:sm:propagation-phase} at \(z=\zeta L\). Its periodicity in \(q\) cancels the dynamical term for uniform full-zone occupation, leaving the Berry polarization at leading order.

For the regular complementary branches, \(\tau(\zeta)=Th(\zeta)\) makes the mode derivative in \eqref{eq:sm:coupled-amplitudes} proportional to \(Th'(\zeta)\):
\begin{equation}
\label{eq:sm:protocol-rate-coupling}
\kpair{v_i}{\partial_\zeta v_{\mathcal B}}
=T h'(\zeta)\,\kpair{v_i}{\partial_\tau v_{\mathcal B}}.
\end{equation}
The first integration by parts in \eqref{eq:sm:oscillatory-ibp} therefore produces a boundary term proportional to \(h'/L\). When \(h'(0)=h'(1)=0\), this contribution vanishes at the entrance and exit. Subtracting the corresponding first adiabatic correction leaves a source of order \(L^{-1}\). A further gap-controlled integration by parts gives exit amplitudes of order \(L^{-2}\) under the stated bounds. The unsigned exit weight is quadratic in these amplitudes, giving
\begin{equation}
\label{eq:sm:endpoint-flat-leakage}
\bm c_\perp(1)=O(L^{-2}),
\qquad
W_\perp(L)=O(L^{-4}).
\end{equation}
For the generally nonnormal Euclidean generator, the estimate uses a uniform reduced-resolvent bound at fixed truncation. Interior amplitudes remain \(O(L^{-1})\). Main-text Fig.~3(d) compares \(h_{\rm lin}(\zeta)=\zeta\) and \(h_{\rm bc}(\zeta)=3\zeta^2-2\zeta^3\); \(h_{\rm bc}\) is the main-text protocol. Figure~\ref{fig:s6}(c) shows the center error, whose oscillatory corrections arise from \eqref{eq:sm:oscillatory-ibp} and obey the \(O(L^{-1})\) estimate in \eqref{eq:sm:finite-length-center}.

The Cauchy energy satisfies a cutoff-independent stability bound. For the \(T\)-periodic factor \(a\) of \(A=\ee^{\ii qt}a\), it obeys \(\dd\mathscr E_q/\dd z=\mu\int_0^T(\partial_z\eps)|D_qa|^2\dd t\) with \(|\partial_z\eps|=O(L^{-1})\). Positivity of \(\eps\) gives \(\dd\mathscr E_q/\dd z\le b(z)\mathscr E_q\), where \(b(z)=\sup_{t\in[0,T]}\frac{|\partial_z\eps(z,t)|}{\eps(z,t)}\). Multiplying by the integrating factor \(\exp[-\int_0^z b(s)\,\dd s]\) and integrating gives
\begin{equation}
\label{eq:sm:cauchy-energy-stability}
\mathscr E_q(z)=\int_0^T\big[|\partial_za|^2+\mu\eps\,|D_qa|^2\big]\dd t\;\le\;\ee^{\Gamma}\,\mathscr E_q(0).
\end{equation}
Here \(\Gamma=\int_0^L b(s)\,\dd s\). For the fixed positive pump profile, \(b(z)=O(L^{-1})\), so \(\Gamma\) is bounded independently of \(L\). The bound is also independent of the harmonic cutoff because it follows directly from \eqref{eq:sm:maxwell}.

The lowest forward and backward branches approach \(k=0\) together. Their unit-action eigenvectors contain the factor \(k_0^{-1/2}\), whereas the pair admits a regular two-dimensional Cauchy basis. Let \(u_0\) denote the \(L^2\)-normalized periodic profile of the lowest band. At fixed \(\tau\), we use the basis \(e_1=(u_0,0)^{\mathsf T}\), \(e_2=(0,u_0)^{\mathsf T}\). In this basis, the frozen propagation equation is \(\partial_z\bm b=\mathsf G_0\bm b\), with
\begin{equation}
\label{eq:sm:low-cauchy-generator}
\mathsf G_0(q,\tau)=
\begin{pmatrix}
0&1\\
-k_0^2(q,\tau)&0
\end{pmatrix},
\qquad
\mathsf G_0(0,\tau)=
\begin{pmatrix}
0&1\\
0&0
\end{pmatrix}.
\end{equation}
At \(q=0\), the profile is the constant \(u_0=T^{-1/2}\), and the frozen generator maps \(e_1\) to zero and \(e_2\) to \(e_1\). The second basis vector is therefore a generalized eigenvector, and the two vectors remain linearly independent at the degeneracy. The spectral projector onto the pair and this Cauchy basis extend smoothly across \(q=0\).

For the continuum operator, the lowest branch has the long-wavelength expansion~\cite{SMPachecoEngheta2020}
\begin{equation}
\label{eq:sm:low-band-expansion}
k_0(q,\tau)=|q|\sqrt{\mu\eps_h(\tau)}+O(|q|^3),
\qquad
\eps_h^{-1}=\frac1T\int_0^T\frac{\dd t}{\eps(t;\tau)} .
\end{equation}
At fixed cutoff, the same linear scaling holds with a cutoff-dependent coefficient. The low-band profile is independent of \(\tau\) at \(q=0\), so its \(\tau\) derivative vanishes linearly with \(q\).

The lowest signed branches accumulate a relative dynamical phase determined by their separation \(2k_0\). At the exit, the phases defined in \eqref{eq:sm:coupled-amplitudes} satisfy
\begin{equation}
\label{eq:sm:low-pair-phase-crossover}
\theta_0^+(q,1)-\theta_0^-(q,1)
=2L\int_0^1 k_0\!\left[q,Th(s)\right]\,\dd s
\asymp L|q|.
\end{equation}
For the fixed pump profile, the linear small-\(|q|\) behavior in \eqref{eq:sm:low-band-expansion} gives an order-unity relative phase at \(|q|\sim L^{-1}\) [Fig.~\ref{fig:s6}(a)]. We use signed branches outside this scale and their common Cauchy block inside it. At \(q=0\), homogeneous propagation gives \(b_1(z+\Delta z)=b_1(z)+\Delta z\,b_2(z)\), producing the length factor in the field component.

For \(|q|\ge L^{-1}\) the branch estimates of \eqref{eq:sm:generic-amplitudes} and \eqref{eq:sm:endpoint-flat-leakage} apply unchanged, with the selected-to-block source \(O(\sqrt{|q|})\,h'(\zeta)\) bounded on the zone.

To describe the driven block in the inner region, we expand the periodic field and its physical \(z\) derivative in the \(L^2\)-orthonormal instantaneous profiles \(u_m\) at each \(q\). Let \(\bm b_m\) denote the corresponding coefficient pair, with \(\bm b_0=\bm b\); these coefficients retain the physical propagation phases. Here \(\langle f,g\rangle=\int_0^T f^*(t)g(t)\,\dd t\) is the cell \(L^2\) inner product. In the gauge \(\langle u_0,\partial_\zeta u_0\rangle=0\), projection of \eqref{eq:sm:maxwell} gives
\begin{equation}
\label{eq:sm:driven-low-block}
\begin{aligned}
\partial_\zeta\bm b
&=L\mathsf G_0\bm b-\sum_{m\ne0}\mathcal C_{0m}\bm b_m,\\
\mathcal C_{0m}
&=\langle u_0,\partial_\zeta u_m\rangle
=-T h'(\zeta)\langle\partial_\tau u_0,u_m\rangle .
\end{aligned}
\end{equation}
The second equality follows by differentiating \(\langle u_0,u_m\rangle=0\) and using \(\tau=Th(\zeta)\). Since \(\partial_\tau u_0=O(q)\), every coupling into the low pair contains a factor \(q\). For each fixed \(L\), this factor and the vanishing initial low-band component give \(\bm b=q\widetilde{\bm b}\). The sum includes the selected band and all other complementary modes.

Here \(b_1\) and \(b_2\) are the field and physical-derivative amplitudes in the lowest Cauchy block. Write the forcing in \eqref{eq:sm:driven-low-block} as \(\bm f_0=(f_1,f_2)^{\mathsf T}=-\sum_{m\ne0}\mathcal C_{0m}\bm b_m\). For \(|q|\le L^{-1}\), introduce \(\widehat{\bm b}=(b_1,Lb_2)^{\mathsf T}\). The driven block then takes the form
\begin{equation}
\label{eq:sm:scaled-low-block}
\partial_\zeta\widehat{\bm b}
=
\begin{pmatrix}
0&1\\
-L^2k_0^2&0
\end{pmatrix}
\widehat{\bm b}
+
\begin{pmatrix}
f_1\\
Lf_2
\end{pmatrix}.
\end{equation}
The linear small-\(|q|\) behavior of \(k_0\) bounds the homogeneous matrix uniformly in the inner region. Its propagator is therefore bounded on \(0\le\zeta\le1\), with a bound independent of \(L\). We estimate the forcing integral with this uniformly bounded propagator.

Let \(\bm a_R\) and \(\bm\beta_R\) collect the field and physical-\(z\)-derivative coefficients of all modes outside the lowest pair, including the selected band. In a smooth orthonormal frame for this regular subspace, denote the connection blocks by \(\mathsf C=\mathcal C_{0R}\), \(\mathsf D=\mathcal C_{RR}\), and \(\mathsf E=\mathcal C_{R0}\). Primes below denote \(\zeta\) derivatives. Define the corrected pair \(\bm w=(b_1,Lb_2+\mathsf C\bm a_R)^{\mathsf T}\). Combining the field and derivative equations gives
\begin{equation}
\label{eq:sm:corrected-low-block}
\partial_\zeta\bm w
=
\begin{pmatrix}
0&1\\
-L^2k_0^2-\mathsf C\mathsf E&0
\end{pmatrix}\bm w
+
\begin{pmatrix}
-2\mathsf C\\
\mathsf C'-\mathsf C\mathsf D
\end{pmatrix}\bm a_R .
\end{equation}
The terms \(-L\mathsf C\bm\beta_R\) and \(+L\mathsf C\bm\beta_R\) cancel in the second component. At fixed cutoff, smoothness of the frame and \(\partial_\tau u_0=O(q)\) give \(\mathsf C,\mathsf C',\mathsf C''=O(|q|)\), \(\mathsf E=O(|q|)\), and bounded \(\mathsf D,\mathsf D'\). Let \(\mathsf K_R\) be the matrix of \(\mathcal L_q\) restricted to the regular subspace. In the neighborhood considered here it has a uniform positive lower bound, and \(\mathsf K_R^{-1}\) and its \(\zeta\) derivative are bounded. The Cauchy-energy estimate~\eqref{eq:sm:cauchy-energy-stability} therefore bounds \(\bm a_R\), \(\bm\beta_R\), and \(b_2\) independently of \(L\). The vanishing initial low component gives \(\bm w(0)=(0,\mathsf C(0)\bm a_R(0))^{\mathsf T}=O(|q|)\). The bounded homogeneous propagator and the \(O(|q|)\) forcing in~\eqref{eq:sm:corrected-low-block} then give \(\bm w=O(|q|)\), and hence \(\widehat{\bm b}=O(|q|)\), throughout the inner region.

When \(h'(0)=h'(1)=0\), the connection block obeys \(\mathsf C(0)=\mathsf C(1)=0\), so \(\bm w(0)=0\) and \(\bm w(1)=\widehat{\bm b}(1)\). The regular physical-derivative equation can be written as
\begin{equation}
\label{eq:sm:regular-field-identity}
\bm a_R
=-\frac1L\mathsf K_R^{-1}
\left(
\partial_\zeta\bm\beta_R+\mathsf D\bm\beta_R+\mathsf E b_2
\right).
\end{equation}
Substituting~\eqref{eq:sm:regular-field-identity} into the driven integral for \(\bm w\) and integrating the \(\partial_\zeta\bm\beta_R\) term by parts gives \(\bm w(1)=O(|q|/L)\). The boundary terms and the remaining integrals have an explicit factor \(L^{-1}\); the coefficient bounds above, together with~\eqref{eq:sm:cauchy-energy-stability}, bound the remaining factors by \(O(|q|)\). Thus \(\widehat{\bm b}(1)=O(|q|/L)\).

The unsigned action weight of the low pair is equivalent to
\begin{equation}
\label{eq:sm:low-block-weight}
W_0=\frac{1}{2\mu}\left(k_0|b_1|^2+\frac{|b_2|^2}{k_0}\right).
\end{equation}
For \(|q|\le L^{-1}\), the low-block equations give \(b_1=O(|q|)\) and \(b_2=O(|q|/L)\) for a generic protocol; when \(h'(0)=h'(1)=0\), the exit values improve to \(b_1=O(|q|/L)\) and \(b_2=O(|q|/L^2)\). Using the linear small-\(|q|\) behavior of \(k_0\), integration of both terms of \eqref{eq:sm:low-block-weight} over \(|q|\le L^{-1}\) gives an inner-region weight \(O(L^{-4})\) for a generic protocol and \(O(L^{-6})\) at a boundary-cancelling exit. These contributions are subleading to the outer-region bounds, so \eqref{eq:sm:generic-amplitudes}, \eqref{eq:sm:finite-length-center}, and \eqref{eq:sm:endpoint-flat-leakage} remain valid, with \(W_\perp(L)=O(L^{-2})\) and \(O(L^{-4})\), respectively.

For the moment derivative, set \(p=|q|/q_0\) with fixed \(q_0>0\), write \(k_0=p\nu\), and rescale \(y_0^\sigma=c_0^\sigma/\sqrt p\). The estimate \(\partial_\zeta u_0=O(q)\) gives regular--low couplings of order \(\sqrt p\). After rescaling, the couplings from the regular modes to the lowest pair are \(O(1)\), while the reverse couplings are \(O(p)\). Their one-sided \(q\) derivatives remain bounded. Applying the \(q\)-differentiated amplitude estimate~\eqref{eq:sm:moment-derivative-bounds}, with the selected-to-complement gap \eqref{eq:sm:signed-gap}, gives \(y_0^\sigma=O(L^{-1})\) and \(\partial_qy_0^\sigma=O(1)\). Writing \(c_0^\sigma=\sqrt p\,y_0^\sigma\) pairs the rescaled amplitudes with the regular vectors \(\widehat v_0^\sigma=\sqrt p\,v_0^\sigma\), preserving each branch contribution. For \(q\ne0\), after removing the common selected-band phase, the low contribution to the remainder in~\eqref{eq:sm:state-decomposition} is
\begin{equation}
\label{eq:sm:low-remainder-reconstruction}
\begin{aligned}
r_0
&=\sum_{\sigma=\pm1}y_0^\sigma
\ee^{\ii(\theta_0^\sigma-\theta_{\mathcal B})}\,
\widehat v_0^\sigma,\\
\widehat v_0^\sigma
&\equiv\sqrt p\,v_0^\sigma
=\sqrt{\frac{\mu}{\nu}}
\begin{pmatrix}
u_0\\
\ii\sigma p\nu u_0
\end{pmatrix}.
\end{aligned}
\end{equation}
Here \(\theta_{\mathcal B}=\theta(q,L\zeta)\), and \(\theta_0^\sigma\) is the dynamical phase of branch \((0,\sigma)\) defined in~\eqref{eq:sm:coupled-amplitudes}. The vectors \(\widehat v_0^\sigma\) and their one-sided \(q\) derivatives are bounded. Using \(y_0^\sigma=O(L^{-1})\) and \(\partial_qy_0^\sigma=O(1)\), differentiation of the amplitudes gives an \(O(1)\) contribution. Differentiation of the relative phases gives another \(O(1)\) contribution, since \(\partial_q(\theta_0^\sigma-\theta_{\mathcal B})=O(L)\). Differentiation of the rescaled vectors contributes \(O(L^{-1})\). Together these estimates give \(\|r_0\|_{\mathrm C}=O(L^{-1})\) and \(\|\partial_qr_0\|_{\mathrm C}=O(1)\). At \(q=0\), the low subspace is invariant and initially unoccupied, so its physical component remains zero. Smooth matching across \(q=0\) follows, at each fixed \(L\), from the full Cauchy evolution and the smooth projector onto the lowest pair.

\end{document}